\documentclass[useAMS,usenatbib,fleqn]{mn2e}
\usepackage{amsmath}
\usepackage{graphicx}
\usepackage{amssymb}
\def\tfrac#1#2{{\textstyle\frac{#1}{#2}}}
\def\nd#1#2{\frac{d #1}{d #2}}

\title[Effect of an expanding universe on masses]{The effect of an expanding universe on massive objects}
\author[Roshina Nandra et al.]{Roshina Nandra$^{1,2}$\thanks{E-mail:
rn288@mrao.cam.ac.uk (RN); a.n.lasenby@mrao.cam.ac.uk (ANL), mph@mrao.cam.ac.uk (MPH)}, Anthony N.
Lasenby$^{1,2}$\footnotemark[1] and Michael P. Hobson$^{1}$\footnotemark[1] \\
$^{1}$Astrophysics Group, Cavendish Laboratory, JJ Thomson Avenue, Cambridge CB3 0HE, U.K.\\
$^{2}$Kavli Institute for Cosmology, c/o Institute of Astronomy, Madingley Road, Cambridge CB3 0HA, U.K.}

\begin{document}

\date{Accepted ---. Received ---; in original form \today}

\pagerange{\pageref{firstpage}--\pageref{lastpage}} \pubyear{2011}

\maketitle

\label{firstpage}

\begin{abstract}
We present some astrophysical consequences of the metric for a point
mass in an expanding universe derived in Nandra, Lasenby \& Hobson,
and of the associated invariant expression for the force required to
keep a test particle at rest relative to the central mass. We focus on
the effect of an expanding universe on massive objects on the scale of
galaxies and clusters. Using Newtonian and general-relativistic
approaches, we identify two important time-dependent physical radii
for such objects when the cosmological expansion is accelerating;
these radii are found to be insensitive to relativistic effects. The
first radius, $r_F$, is that at which the total radial force on a test
particle is zero, which is also the radius of the largest possible
circular orbit about the central mass $m$ and where the gas pressure
and its gradient vanish. The second radius, $r_S$, is that of the
largest possible stable circular orbit, which we interpret as the
theoretical maximum size for an object of mass $m$. The radius $r_S$
is typically smaller than $r_F$ by a factor $\sim 1.6$.  In contrast,
for a decelerating cosmological expansion, no such finite radii
exist. Assuming a cosmological expansion consistent with a
$\Lambda$CDM concordance model, at the present epoch we find that these radii put a sensible constraint on the typical
sizes of both galaxies and
clusters at low redshift. For galaxies, we also find that these radii agree closely with zeroes in the radial
velocity field in the neighbourhood of nearby galaxies, as inferred by
Peirani \& Pacheco from recent observations of stellar velocities. We
then consider the future effect on massive objects of an accelerating
cosmological expansion driven by phantom energy, for which the
universe is predicted to end in a `Big Rip' at a finite time in the
future at which the scale factor and the Hubble parameter become
singular. In particular, we present a novel way of calculating the
time prior to the Big Rip that an object of a given mass and size will
become gravitationally unbound.
\end{abstract}


\section{Introduction}

In \citet{NandraPub1} (hereafter NLH1), we considered the effect of a
massive object on an expanding universe by deriving the metrics that
describe a point mass in spatially-flat and spatially-curved
cosmologies. In particular, we derived a general invariant expression
for the force required to keep a test particle at rest relative to the
point mass.  Conversely, one would expect cosmological expansion to
have a significant effect on the dynamics of massive objects.  In this
paper, we investigate this effect for objects on the scale of galaxies
and clusters.

Considering the radial motion of a test particle in a spatially-flat expanding universe, we showed in NLH1 that, in
the Newtonian limit, the radial force $F$ per unit mass at a distance $r$ from a point mass $m$ is given by
\begin{equation}
F = -\frac{m}{r^2} - q(t)H^2(t)r,
\label{eq:eq1}
\end{equation}
where $H(t) \equiv R'(t)/R(t)$ is the Hubble parameter, $q(t) =
-H'(t)/H^2(t)-1$ is the deceleration parameter, $R(t)$ is the
universal scale factor and primes denote derivatives with respect to
the cosmic time $t$.\footnote{We will
  work in natural units throughout, so that $c=G=1$.} Thus, the force
consists of the usual time-independent $1/r^2$ component due to the
central mass and a time-dependent cosmological component proportional
to $r$ that is directed outwards (inwards) when the cosmological
expansion is accelerating (decelerating); see also \cite{davis}.  For
the currently favoured spatially-flat $\Lambda$CDM cosmology, the
cosmological force reverses direction at redshift $z \approx 0.67$.

We will consider this general case (and its general-relativistic
extension) in due course, but the main physical effects of the
present-day accelerating cosmological expansion on the dynamics of
massive objects can be illustrated by considering the special case of
a de Sitter universe, which contains no cosmological matter fluid (or
radiation), but only dark energy in the form of a non-zero
cosmological constant $\Lambda$.  In this case, the Hubble parameter
and, hence, the deceleration parameter become time-independent and are
given by $H=\sqrt{\Lambda/3}$ and $q=-1$. Thus, the force
(\ref{eq:eq1}) also becomes time-independent and reads
\begin{equation}
F = -\frac{m}{r^2} + \tfrac{1}{3}\Lambda r.
\label{eq:eq2}
\end{equation}

The time-independence of this expression is related to the well-known
(but curious) fact that the presence of a pure non-zero cosmological
constant can be treated mathematically in a completely static
manner. As we will discuss in Section~\ref{sec:expansion}, even in the
full general-relativistic case, the metric for a point mass embedded
in an expanding de Sitter cosmological background is connected by a
simple coordinate transformation to the standard
Schwarzschild--de Sitter metric, which is static. Thus, in this
special case, this correspondence unburdens us from the need to
include directly the expanding cosmological background and allows us
to consider simply the static case of a point mass in the presence of
a non-zero cosmological constant. Indeed, (\ref{eq:eq2}) is easily
obtained by considering directly the Newtonian limit of the Einstein
field equations with $\Lambda \neq 0$ in this case
\citep{ClassMechBook}. Moreover, this correspondence enables us to
generalise straightforwardly to spatially finite (i.e. non-pointlike)
spherically-symmetric massive objects. In the Newtonian limit, for
example, (\ref{eq:eq2}) is replaced simply by
\begin{equation}
F = -\frac{M(r)}{r^2} + \tfrac{1}{3}\Lambda r,
\label{eq:eq3}
\end{equation}
where $M(r)$ is the total mass of the object contained within the
radius $r$. If the object has the radial density profile $\rho(r)$, then
$M(r) = \int_0^r 4\pi \bar{r}^2 \rho(\bar{r})\,d\bar{r}$.

Although the de Sitter background is not an accurate representation of
our universe, the standard cosmological model is dominated by
dark-energy in a form consistent with a simple cosmological constant,
which is driving the current observed acceleration of the universal
expansion \citep{straumann, hubble, Riess, kratochvil, garnavich}. It
is therefore of interest to consider the astrophysical (as opposed to
cosmological) consequences of a non-zero $\Lambda$, which 
has previously been considered only in the context of growth of
structure analyses \citep{lahav91}.

Even in the simple Newtonian case, embodied by the force expression
(\ref{eq:eq3}), we see immediately that there is an obvious, but
profound, difference between the cases $\Lambda=0$ and $\Lambda \neq
0$. In the former, the force on a constituent particle of a galaxy or
cluster (say) is attractive for all values of $r$ and tends gradually
to zero as $r \to \infty$ (for any sensible radial density profile).
In the latter case, however, the force on a constituent particle (or
equivalently its radial acceleration) {\em vanishes} at the {\em
  finite} radius $r_F$ which satisfies $r_F=[3M(r_F)/\Lambda]^{1/3}$,
beyond which the net force becomes repulsive. This suggests that a
non-zero $\Lambda$ should set a {\em maximum size}, dependent on mass, for
galaxies and clusters.

\begin{figure}
\centering
\includegraphics[width=7cm]{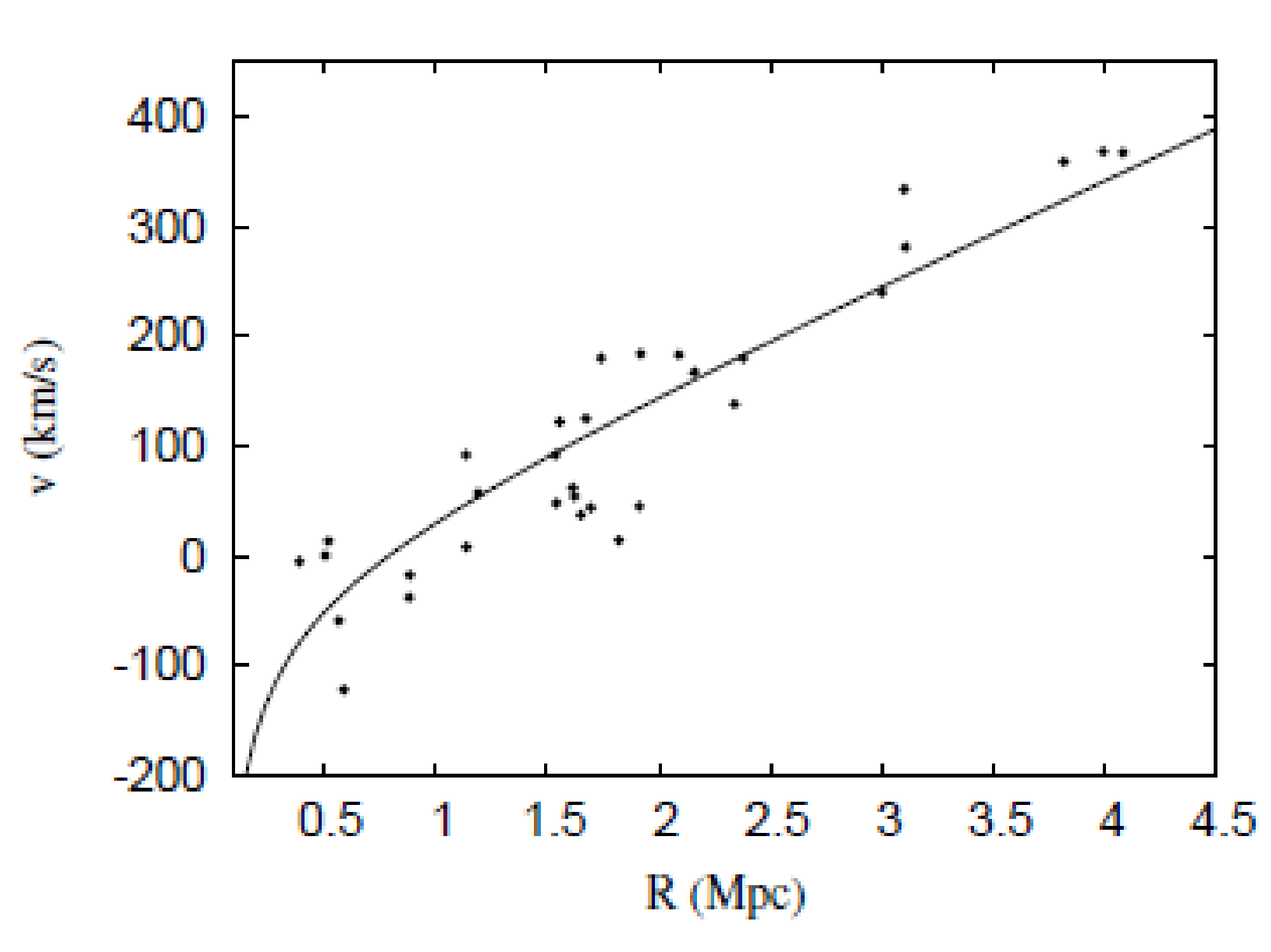}
\caption{Radial-velocity versus distance of stars in the M81
  group, which has a mass of $9.2\times 10^{11}$ M$_\odot$, and the
  best-fit line (\ref{eq:PPbestfit}) \citep[reproduced from][]{peirani1}.}\label{fig:peirpech}
\end{figure}

Some observational evidence for this suggestion has been reported by
\citet{peirani2, peirani1}, who analysed the radial velocities of
stars in the neighbourhood of a number of nearby galaxies, including
M83, M81, IC 342 and NGC 253, by collating a number of unrelated
observational studies \citep{karachentsev}. Note that all these
galaxies lie at very low redshift ($z \sim 10^{-4}$--$10^{-3}$). The
data for M81 are shown in Fig.~\ref{fig:peirpech}. To
obtain the best-fit line for each galaxy, it was modelled as a system
consisting of low-mass satellites sitting within expanding shells
around a dominant core of mass $m$.  Such an approach had
    already been proposed by \citet{LyndenBell2} and
    \citet{sandage} and is based on the Lema\^{i}tre-Tolman (LT) model
    \citep{lemaitre,tolman}.  One essentially begins with the simple
    Newtonian expression (\ref{eq:eq3}) for the radial acceleration
    from which an expression for the radial velocity can be
    calculated, containing initially unknown constants.  Obtaining an
    analytical expression in the case of a non-zero cosmological
    constant is problematic, so Peirani \& Pacheco instead found a
    numerical solution to the relativistic LT model, valid for
    $z\approx 0$:
\begin{equation}
v_{\rm radial}(r)=-\frac{0.976}{r^{n}}\left(\frac{m}{H_{0}^{\ 2}}\right)^{\frac{n+1}{3}}+1.377H_{0}r,\label{eq:PPbestfit}
\end{equation}
where $n=0.627$ and $H_0$ is the present-day Hubble constant. This
relation appears to fit the data extremely well, and highlights the
existence of a well-defined `cut-off' radius, depending on the galaxy
mass, at which the radial velocity is zero, separating bound and
unbound material. We see that this radius is $\sim 0.8$~Mpc for M81.  This is around 3 times the typical extent of $\sim 0.26$~Mpc
quoted in \citet{klypin} for the Milky Way, which has a similar mass to M81.

The numerical analysis of \citet{peirani2, peirani1} is based on radial velocities, and it relies on observational data to fit values to the initially unknown constants.  To avoid this reliance, in this
paper we consider instead test particles in circular orbits about the
central mass, which is also simpler and potentially more
representative of the trajectories of constituent particles of
galaxies and clusters.  From (\ref{eq:eq3}), in the Newtonian limit,
the speed of a particle in a circular orbit of radius $r$ is given
simply by
\begin{equation}
v(r)=\sqrt{\frac{M(r)}{r}-\frac{1}{3}\Lambda r^2},\label{eq:newtcircvels}
\end{equation}
from which it is clear that no circular orbit can exist beyond the
radius $r_F$. A second physically significant radius $r_S$ is that of
the largest {\em stable} circular orbit, which lies somewhat within
$r_F$, as we will discuss in Section~\ref{sec:expansion}. Indeed, the
latter provides a more physically meaningful limit on the maximum
sizes of galaxies and clusters.

The existence and stability of circular orbits around a central mass
in an expanding universe have been considered previously by
\citet{sussman}, \citet{nowakowski1}, \cite{now3} and
\citet{nowakowski2}. Some details of their work are given in
Section~\ref{sec:comparison}. In particular, they assumed a de Sitter
cosmological background and treated the central mass as pointlike. In
this paper, we will relax both these assumptions, although not
simultaneously.

The overall structure of this paper is as follows.  In
Section~\ref{sec:Newtonian}, we perform a circular velocity analysis
based on the simple Newtonian expression (\ref{eq:newtcircvels})
applied to two assumed radial density profiles $\rho(r)$ for the
central object: a decaying exponential and an NFW profile
\citep{navarro,matos}. We compare our results with the radial velocity
analysis of \citet{peirani2, peirani1} and also derive an analytical
expression for the pressure profile. In Section~\ref{sec:DSGR}, we
extend our analysis to the general-relativistic case, for which we
obtain an analytical circular velocity expression in the linearised
case and employ a numerical approach when considering the full
Einstein equations. In the latter case, we again consider the
pressure profile in the central object.  In
Section~\ref{sec:expansion}, we remove our assumption of a de Sitter
background cosmology and consider the currently-favoured $\Lambda$CDM
model with $\Omega_{\rm m,0}=0.3$ and $\Omega_{\Lambda,0}=0.7$, albeit
at the cost of having to model the central object as a point mass. By
comparing with the de Sitter background model in this case we thus
obtain `correction' factors that may be applied to our earlier results
for the radius $r_F$.  We also obtain an expression for the radius
$r_S$ of the largest stable circular orbit. In
Section~\ref{eq:bigrip}, we investigate the more speculative scenario
that the expansion of the universe is driven by phantom energy,
leading to the prediction that the universe will end in a `Big Rip'
\citep{caldwell}.  In particular, we use our force expression to find
the times prior to the Big Rip at which objects of a given mass and
size would be expected to become unbound. Our conclusions are
presented in Section~\ref{sec:conc}.

Finally, we note that in this paper we model
    galaxies/clusters as being composed of a single `phenomenological'
    fluid, with a single overall density profile and a single
    associated (effective) pressure required for stability; this is
    explained in more detail in the next section.  This avoids the
    complexity of an explicit non-linear multi-fluid treatment,
    whereby one would separate the fluid into its baryonic and dark
    matter components.  The analyses we perform in this work are
    therefore designed to highlight general features, and only
    indicate the existence and approximate locations of the radii
    $r_F$ and $r_S$.

\section{Newtonian de Sitter analysis}
\label{sec:Newtonian}

We first adopt the simple Newtonian expression (\ref{eq:newtcircvels})
for the circular velocity.  To plot the circular velocity
    function we must choose a form for the density distribution.  The
    central massive object is modelled as a total `effective' fluid,
    made up of two components: baryonic gas and dark matter. It is
    assumed that both components follow a common density distribution
    (up to an overall normalisation), such that the total density
    $\rho(r) = \rho_{\rm{g}}(r)+\rho_{\rm{dm}}(r)$, where
    $\rho_{\rm{g}}(r)=f_{\rm{g}}\rho_{\rm{dm}}(r)/(1-f_{\rm{g}})$ and
    $f_{\rm{g}} = \rho_{\rm{g}}(r)/\rho(r)$ is the uniform gas
    fraction throughout the object.  Specifically, assuming spherical
    symmetry, we consider two model radial profiles.  The first is a
    decaying exponential
\begin{equation}
\rho(r)=\rho_{0}e^{-\frac{r}{r_{0}}},
\label{eq:expprofile}
\end{equation}
where $r_0$ is a characteristic physical scale of the object and 
$\rho_0$ is its central density. The second is a 
Navarro, Frenk and White (NFW) profile 
\citep{navarro,matos} 
\begin{equation}
\rho(r)=\frac{\rho_{0}r_{0}}{r\left(1+\frac{r}{r_{0}}\right)^2}, 
\label{eq:nfwprofile}
\end{equation}
where $r_0$ is again a characteristic physical scale, but $\rho_0$ is
a characteristic density in this case, since the central density of
(\ref{eq:nfwprofile}) is infinite. In each case, we fix the constant
$r_{0}$ directly and then determine $\rho_0$ by assuming the mass
$M(r_T)$ inside a `typical' size $r_T$
for low-redshift galaxies and clusters
as listed in Table~\ref{table:typicalnos}. We also assume the value of
the cosmological constant to be $\Lambda=10^{-35}$ s$^{-2}$, which is
consistent with observations \citep{carmeli,carroll,weinberg,WMAP}.
\begin{table}
\begin{center}
\begin{tabular}{@{}ccc@{}}\hline
Quantity & Galaxies & Clusters \\ \hline
$r_0$ (kpc) & $3$ & $500$ \\ 
$r_T$ (Mpc) & $0.26$ & $5$ \\ 
$M(r_T)$ ($\text{M}_\odot$) & $10^{12}$ & $10^{15}$ \\ \hline 
\end{tabular}
\caption{Summary of the values assumed in our modelling of
  galaxies and clusters.  Note that the values for galaxies correspond to the virial radius/mass of the Milky Way, as given in Table 2 of \citet{klypin}. \label{table:typicalnos}}
\end{center}
\end{table}

\begin{figure*}
\centering
\begin{tabular}{cc}
\begin{minipage}{2.5in}
\centering
\fbox{\includegraphics[height=1.6in,width=2.2in]
{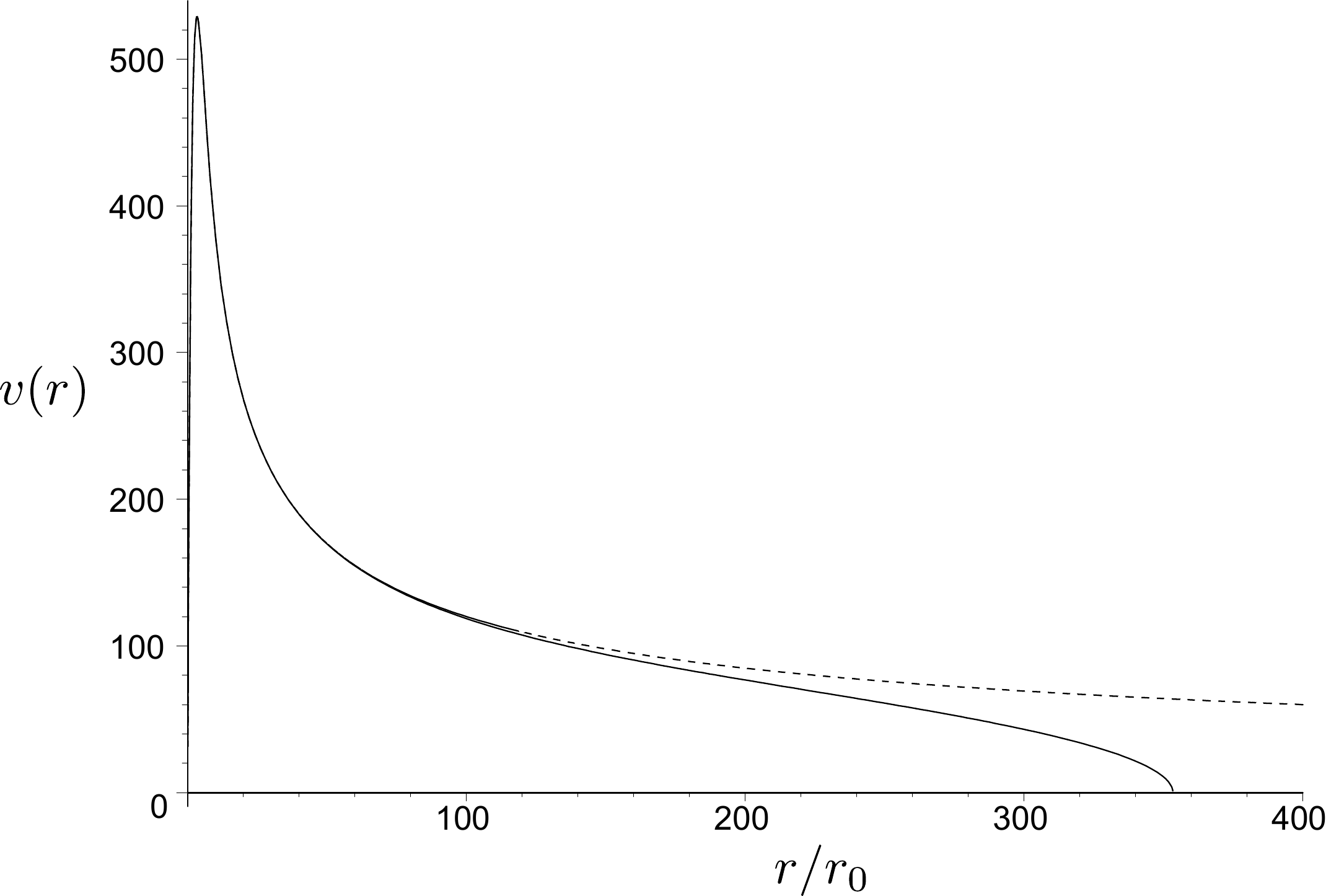}}
\end{minipage}
&
\begin{minipage}{2.5in}
\centering
\fbox{\includegraphics[height=1.6in,width=2.2in]
{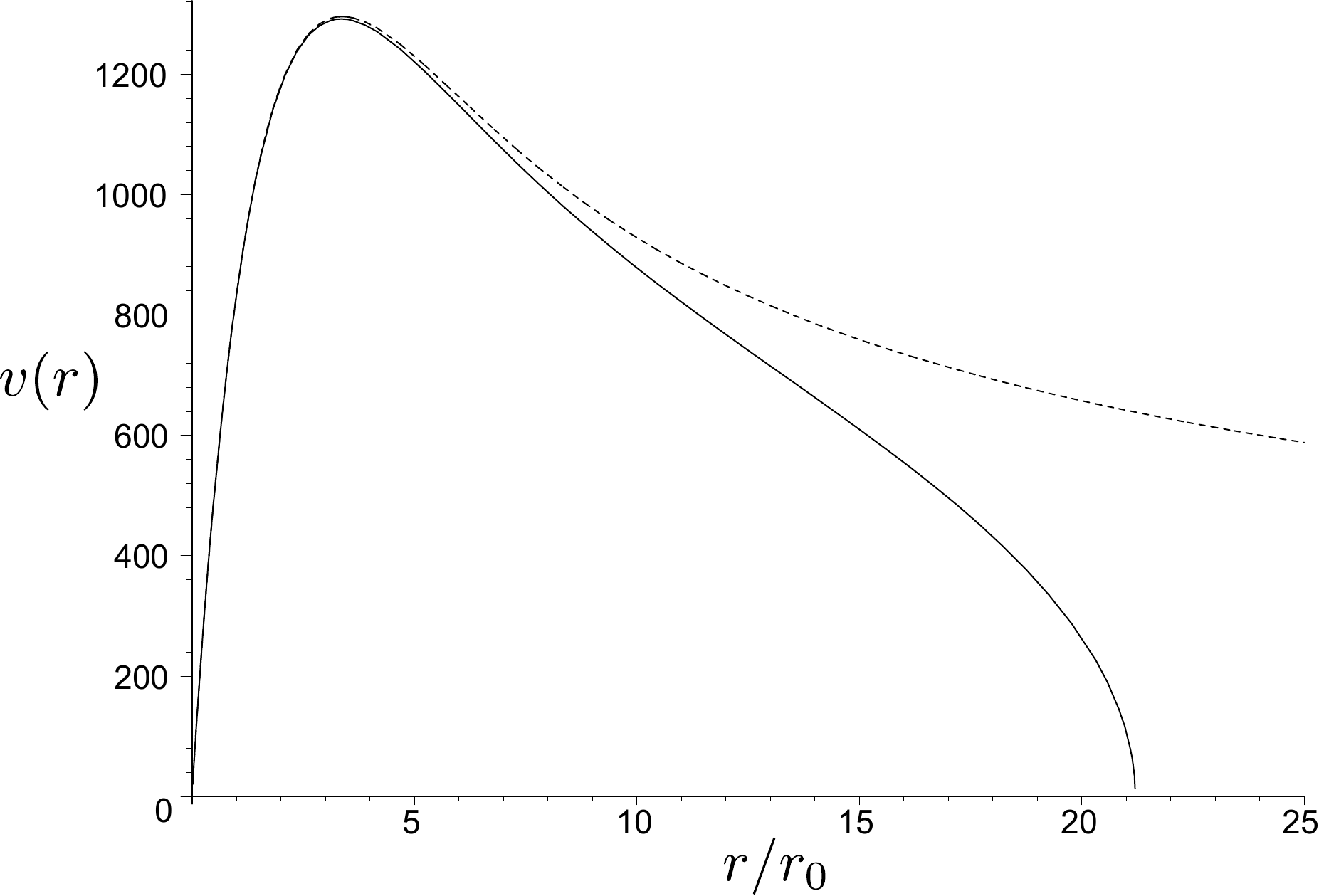}}
\end{minipage}
\\
\\
(a) Galaxy with exponential $\rho(r)$ & 
(b) Cluster with exponential $\rho(r)$
\\
\newline
\\
\begin{minipage}{2.5in}
\centering
\fbox{\includegraphics[height=1.6in,width=2.2in]
{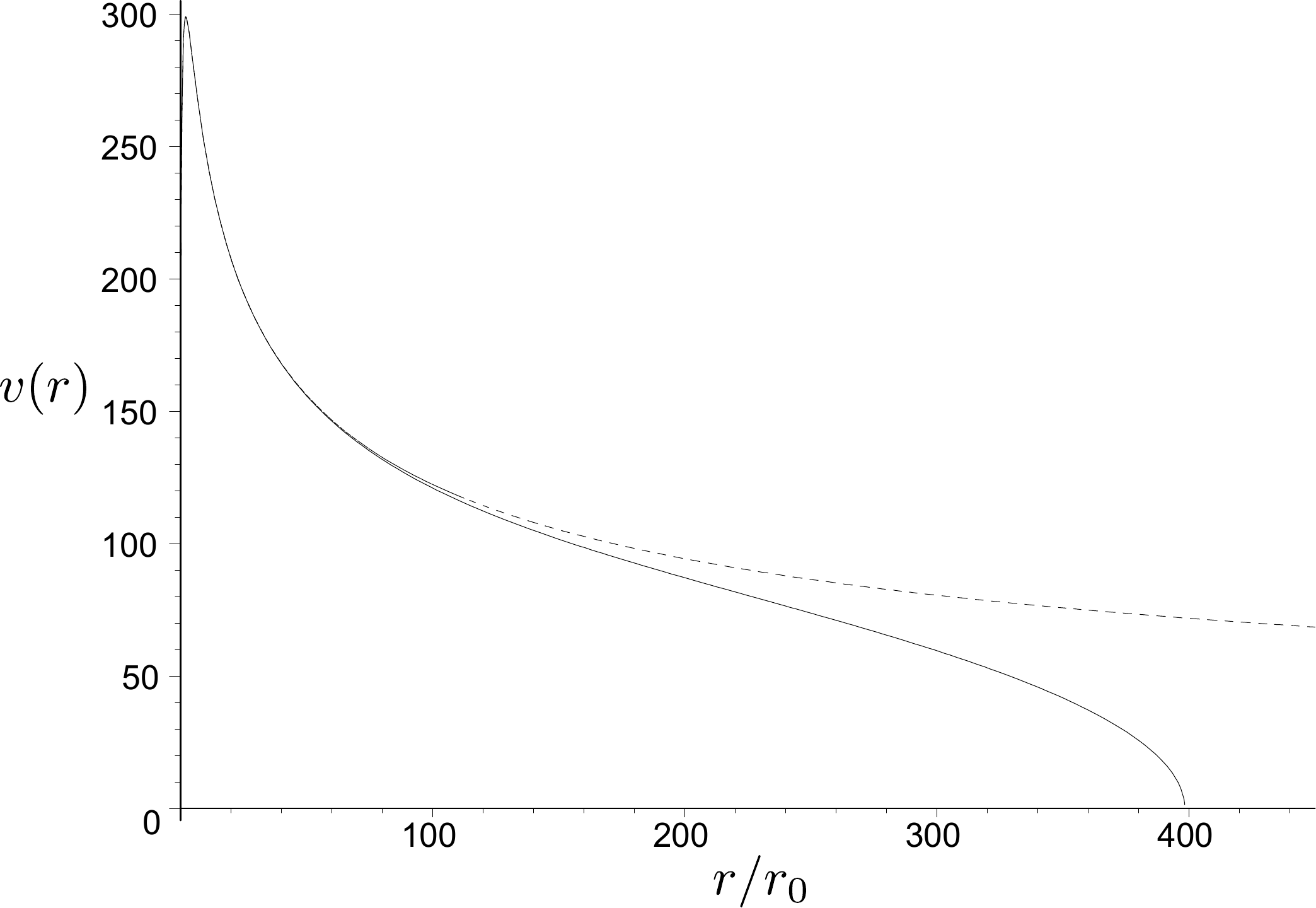}}
\end{minipage}
&
\begin{minipage}{2.5in}
\centering
\fbox{\includegraphics[height=1.6in,width=2.2in]
{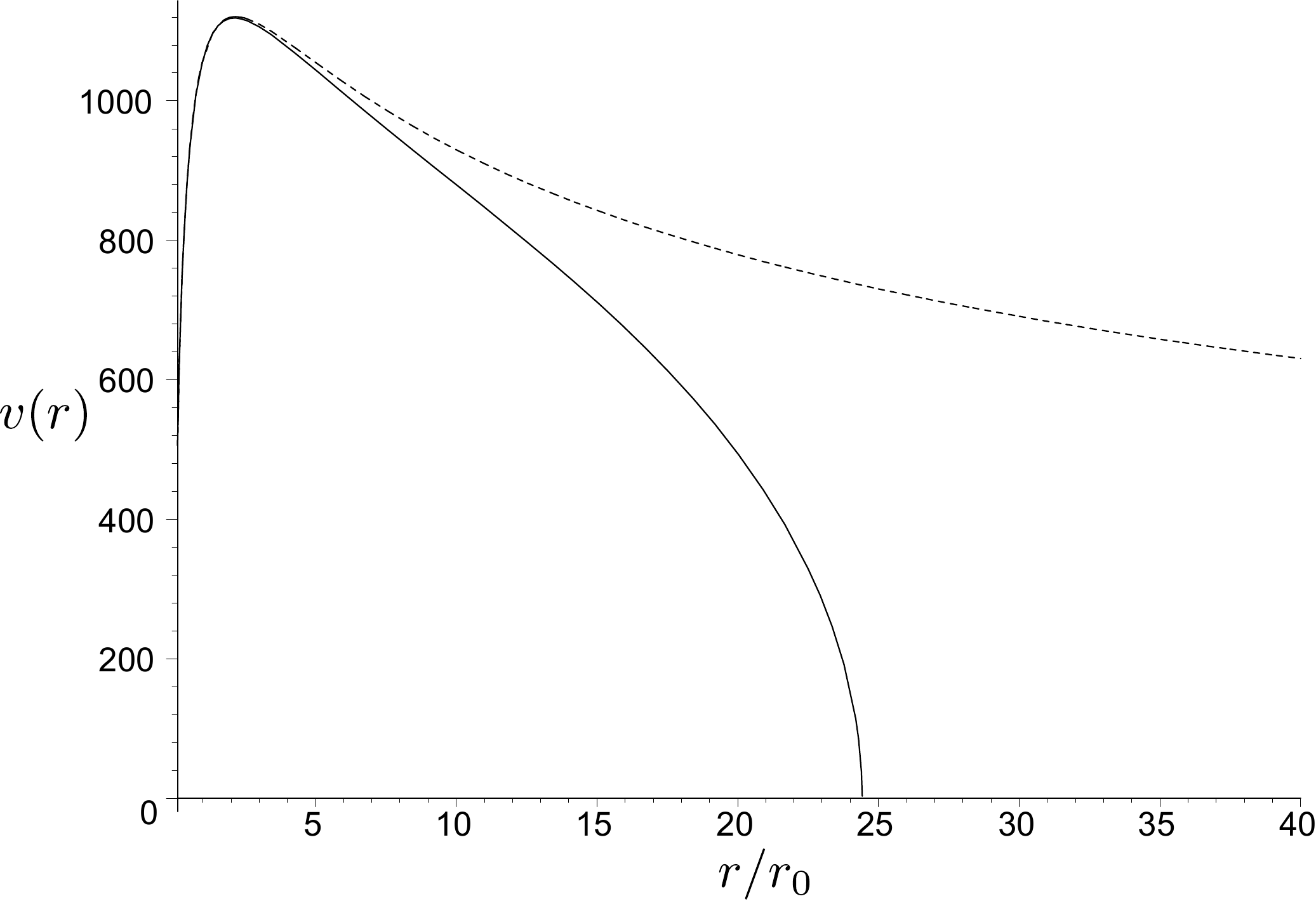}}
\end{minipage}
\\
\newline
\\
(c) Galaxy with NFW $\rho(r)$ &
(d) Cluster with NFW $\rho(r)$
\end{tabular}
\caption{Circular velocity $v(r)$, in km/s, 
obtained from the Newtonian expression (\ref{eq:newtcircvels})
assuming $\Lambda=10^{-35}$ s$^{-2}$ (solid line) and
$\Lambda=0$ (dotted line).} \label{fig:newtonian}
\end{figure*}

The resulting Newtonian circular velocity (\ref{eq:newtcircvels}) is
plotted in Fig.~\ref{fig:newtonian}.
As expected, in each case the circular velocity, and hence the radial
force, falls to zero at a finite radial distance from the centre of
the object.  This is in sharp contrast to the situation when $\Lambda
=0$ (also plotted in Fig.~\ref{fig:newtonian}), in which case there is
no definite cut-off radius. Although, for $\Lambda \neq 0$, the radial
force vanishes at $r_F$, this marks the position of an unstable
equilibrium, since any radial displacement of a test particle would
move it to a region where either the gravitational or cosmological
force dominates \citep{faraoni}.  We note that for the galaxy and the
cluster, the cut-off radius $r_F$ is smaller for the exponential
radial density profile than for the NFW profile.  Table
\ref{table:factors} gives the cut-off radii $r_F$ for the modelled
objects.

One may also use the plots in Fig. \ref{fig:newtonian} to
    read off the maximum circular velocities $v_{\rm{max}}$, which
    clearly lie close to the centres of the modelled objects.  Using a decaying exponential profile (\ref{eq:expprofile}), for a typical galaxy $v_{\rm{max}}\sim530$~km/s, and for a typical
    cluster $v_{\rm{max}}\sim1280$~km/s.  Since the total effective fluid is largely dominated by dark matter, an NFW profile (\ref{eq:nfwprofile}) is likely to be more realistic.  This leads to $v_{\rm{max}}\sim300$~km/s for a typical galaxy and $v_{\rm{max}}\sim1120$~km/s for a typical cluster.  Indeed in this case the maximum circular velocity for a galaxy is only about $1.3$ times higher than that found by \citet{klypin}.

\begin{table}
\begin{center}
\begin{tabular}{@{}lcc@{}}\hline
 & Galaxies & Clusters \\ \hline
Exponential $\rho(r)$ & $1.06$ & $10.5$ \\ 
NFW $\rho(r)$ & $1.20$ & $12.0$ \\ \hline
\end{tabular}
\end{center}
\caption{The distance $r_F$ (in Mpc) at which the total radial force and circular velocities vanish for `galaxies' and `clusters', using a Newtonian de Sitter analysis and the values for parameters given in Table \ref{table:typicalnos}.  Note that the $r_F$ values are read from the plots in Fig. \ref{fig:newtonian}.}\label{table:factors}
\end{table}

Having laid down two possible density profiles, we may find
    the associated pressure $p(r)$ required to keep the overall fluid
    stable.  This too is made up of two components: an ordinary
    pressure associated with the gas $p_{\rm{g}}(r)$ and an
    effective pressure associated with the dark matter
    $p_{\rm{dm}}^{\rm{eff}}(r)$.  The latter arises when the
    motions of dark matter particles have suffered phase-mixing and
    relaxation, such that their velocities are randomised and the
    object is virialised, as explained by \citet{LyndenBell} and more
    recently by \citet{binney}.  The total pressure is then given by
    $p(r)=p_{\rm{g}}(r)+p_{\rm{dm}}^{\rm{eff}}(r)$.

The controlling equation for gas in hydrostatic equilibrium is
\begin{equation}
\frac{d}{dr}(p_{\rm{g}}(r))=-\rho_{\rm{g}}(r)\frac{d\Phi(r)}{dr},\label{eq:gaspressure}
\end{equation}
where $\Phi(r)$ is the gravitational potential.  The corresponding
equation for the dark matter, modelled as composed of collisionless
particles which have achieved dynamical equilibrium, is the Jeans
equation,
\begin{equation}
\frac{d}{dr}(\rho_{\rm{dm}}(r)\sigma_r^{2}(r))=-\rho_{\rm{dm}}(r)\frac{d\Phi(r)}{dr},\label{eq:DMpressure}
\end{equation}
where $\sigma_r^{2}(r)$ is the radial velocity dispersion
\citep{tormen}.  Note that we have ignored a possible `softening'
correction by assuming an isotropic velocity distribution,
which follows naturally from our assumption of full spherical
symmetry.  Comparing (\ref{eq:DMpressure}) with
(\ref{eq:gaspressure}), one can see that the two equations are in
identical form, and therefore that the effective pressure of the dark
matter is given by
\begin{equation}
p_{\rm{dm}}^{\rm{eff}}(r)=\rho_{\rm{dm}}(r)\sigma_r^{2}(r).\label{eq:sigmarel}
\end{equation}

We assume that the only interaction between the gas and the dark
matter is gravitational, so $\Phi(r)$ corresponds to the total
gravitational potential.  In the presence of a non-zero cosmological
constant, for a fixed total mass $M(r)$ within a radius $r$, the
gravitational potential is $\Phi(r)=-\int^{\infty}_{r}\left(M(r^{\prime})/r^{\prime2}\right)dr^{\prime}-(1/6)\Lambda
r^2$.  Thus, adding the two equations (\ref{eq:gaspressure}) and
(\ref{eq:DMpressure}) together leads to the standard Newtonian
equation for hydrostatic equilibrium, namely  
\begin{equation}
\frac{dp(r)}{dr}=-\frac{\rho(r)[M(r)-\frac{1}{3}\Lambda r^3]}{r^2}.
\label{eq:OVNewtLambda}
\end{equation}
For $\Lambda=0$, we see immediately that $dp(r)/dr$ is always
negative, tending to zero as $r \to \infty$. For $\Lambda \neq 0$,
however, the behaviour is radically different and using
(\ref{eq:newtcircvels}) we see that the pressure profile has a minimum
precisely at $r=r_F$.

For brevity, we will consider only the (possibly more realistic) NFW
profile.  With an appropriate choice of boundary condition, one may in
fact obtain an analytical form for $p(r)$ (albeit rather unwieldy) for
general $\Lambda$, which we believe has not been reported
previously. The boundary condition in both the $\Lambda=0$
and $\Lambda \neq 0$ cases is that the pressure itself vanishes at the
point where $dp(r)/dr=0$. Thus, for $\Lambda = 0$, one has $p(r) \to
0$ as $r \to \infty$, whereas for $\Lambda \neq 0$ one requires
$p(r_F)=0$. In the latter case, one thus has yet another physically
important interpretation of the radius $r=r_F$. One finds that the
pressure profile is given by
\begin{align}
p(r)=&
-\frac{4\pi\rho_0^2r_0^2\ln\left(1+\frac{r}{r_0}\right)}{1+\frac{r}{r_0}}-\frac{12\pi\rho_0^2r_0^2}{1+\frac{r}{r_0}}\nonumber\\
&+12\pi\rho_0^2r_0^2\text{dilog}\left(1+\frac{r}{r_0}\right)+6\pi\rho_0^2r_0^2\left[\ln\left(1+\frac{r}{r_0}\right)\right]^2\nonumber\\
&-2\pi\rho_0^2r_0^2\ln\left(\frac{r}{r_0}\right)-\frac{2\pi\rho_0^2r_0^3}{r}+\frac{2\pi\rho_0^2r_0^4\ln\left(1+\frac{r}{r_0}\right)}{r^2}\nonumber\\
&-\frac{8\pi\rho_0^2r_0^3\ln\left(1+\frac{r}{r_0}\right)}{r}+2\pi\rho_0^2r_0^2\ln\left(1+\frac{r}{r_0}\right)\nonumber\\
&-\frac{2\pi\rho_0^2r_0^2}{\left(1+\frac{r}{r_0}\right)^2}
-\frac{\Lambda\rho_0 r_0^2}{3\left(1+\frac{r}{r_0}\right)} + C,
\label{eq:analt}
\end{align}
where $\text{dilog}(1+r)=\int^{1+r}_{0}\frac{\ln(t)}{1-t}dt$; it is
worth noting that $\text{dilog}(1+r) \to -\frac{1}{2}[\ln r]^2 -
\frac{1}{6} \pi^2$ as $r \to \infty$. The constant of integration $C$
is set by the above boundary condition on $p(r)$. For the $\Lambda=0$
case, it has the analytical value $C=2\pi^3r_0^2\rho_0^2$, whereas for
$\Lambda\neq 0$ its value was found numerically.

Rather than the pressure profile $p(r)$ itself, of more
    interest, perhaps, is the ratio $p(r)/\rho(r)$.  Assuming a
    perfect gas law and mean molecular weight of $m_{\rm p}/2$, we can
    relate this numerically to a temperature, in K, via
\begin{equation}
T(r) = \frac{m_{\rm p}}{2k_{\rm B}}\left[\frac{p(r)} {\rho(r)}\right]_{\rm{SI}} = \frac{m_{\rm p} c^2}{2k_{\rm B} }\left[\frac{p(r)}{\rho(r)}\right]_{\rm{nat}},\label{eq:gastemp}
\end{equation}
where $m_{\rm p}$ is the proton mass, $k_{\rm B}$ is Boltzmann's
constant and the ratio $p(r)/\rho(r)$ may be expressed in Nm/kg (SI
units), or in natural units which we mostly adopt in this paper.  This
temperature profile is, in fact, that of the gas,
$T(r)=T_{\rm{g}}(r)$. This follows immediately from our assumption
that the gas fraction $f_{\rm{g}}$ is uniform throughout the
galaxy/cluster, in which case, using (\ref{eq:gaspressure}) and
(\ref{eq:DMpressure}) respectively, one has
\begin{equation}
\frac{p}{\rho}=\frac{p_{\rm{g}}}{\rho_{\rm{g}}}=\frac{p_{\rm{dm}}^{\rm{eff}}}{\rho_{\rm{dm}}}.\label{eq:prhoequivalence}
\end{equation}

For $\Lambda=0$, the temperature profile (\ref{eq:gastemp}) can be expressed in a universal
scale-free form as
\begin{align}
T(x) = & \frac{\pi G \rho_0 r_0^2 m_{\rm p}}{k_{\rm B}x}
\left\{(x^3-5x^2-3x+1)(1+x)\ln(1+x)\right. \nonumber \\
& + x^2(1+x)^2[6\,{\rm dilog}(1+x) + 3(\ln (1+x))^2-\ln x] \nonumber \\
& + \left. x[\pi^2 x^3+(2\pi^2-7)x^2+(\pi^2-9)x-1]\right\},
\label{eq:unitemp}
\end{align}
where $x \equiv r/r_0$. However, for $\Lambda \neq 0$, the term in
$\Lambda$ in (\ref{eq:analt}) breaks scale invariance, and we can no
longer write the temperature profile in this universal way.

The peak in the temperature profile (\ref{eq:unitemp}) occurs at $x =
x_m \approx 0.76$, and in terms of $x_m$ one can show analytically
that the peak temperature has the value
\begin{equation}
T_{\rm max} = \frac{x_T(1+x_T)\left[(1+x_m)\ln(1+x_m)-x_m\right]}
{2x_m(1+3x_m)\left[(1+x_T)\ln(1+x_T)-x_T\right]} \frac{GM_T m_{\rm p}}{r_T
k_{\rm B}},\label{eq:peaktemptemp}
\end{equation}
where, as above, $M_T$ is the object's mass at radius $r_T$, and
$x_T=r_T/r_0$.  Since the total gravitational potential energy of the object
is $\sim GM_T^2/r_T$, and the kinetic energy of the constituent
particles is $\sim (M_T/m_p)k_B T$, it follows that our expression for
$T_{\rm max}$ (\ref{eq:peaktemptemp}) in fact corresponds to an analytical
factor times the the virial temperature.  For the `typical' values of
$r_T$ and $r_0$ for a
cluster given in Table~\ref{table:typicalnos}, $x_T=10$ and we obtain
\begin{equation}
T_{\rm max} \approx 0.32\frac{GM_T m_{\rm p}}{r_T k_{\rm B}} \approx 3.3
\times 10^7 {\rm \, K}\label{eq:peaktemp}
\end{equation}
or $T_{\rm max}\sim 2.8 {\rm \, keV}$.  We may compare this
calculated value against the virial temperatures of similar sized clusters as inferred from X-ray data and simulations; indeed our value for $T_{\rm max}$ is in
     close agreement with the simulations of \citet{roncarelli} and lies within the range found in X-ray observations by \citet{vikhlinin}.
     
       Note that for
      a galaxy with the parameters in Table 1 we can calculate that
      $T_{\text{max}}\approx 2.4\times10^6{\rm \, K}$. There has been
long dispute about whether a component of hot halo gas exists for galaxies, but recent X-ray observations,
summarised in \citet{crain}, do support such a component.  In fact
the temperature of the hot gas halo of the Milky Way itself (the mass
and size parameters of which were used in Table 1) has recently been
measured \citep{henley} and lies in the range $1.8-2.4 \times 10^6
{\rm \, K}$, in good agreement with our calculated figure. Of course the
behaviour of the gas in the inner regions of galactic haloes and
clusters is  more or less independent of the effects of $\Lambda$ or
universe expansion (see Fig. 3, where the effects of $\Lambda$ only
become perceptible to the right of the peak in temperature), but we
mention these details, and corroborating observations, to show that our
model does work sensibly in these inner regions, and therefore plausibly
reflects reality in the outer regions as well.

\begin{figure*}
\centering
\begin{tabular}{cc}
\begin{minipage}{2.5in}
\centering
\fbox{\includegraphics[height=1.6in,width=2.5in]
{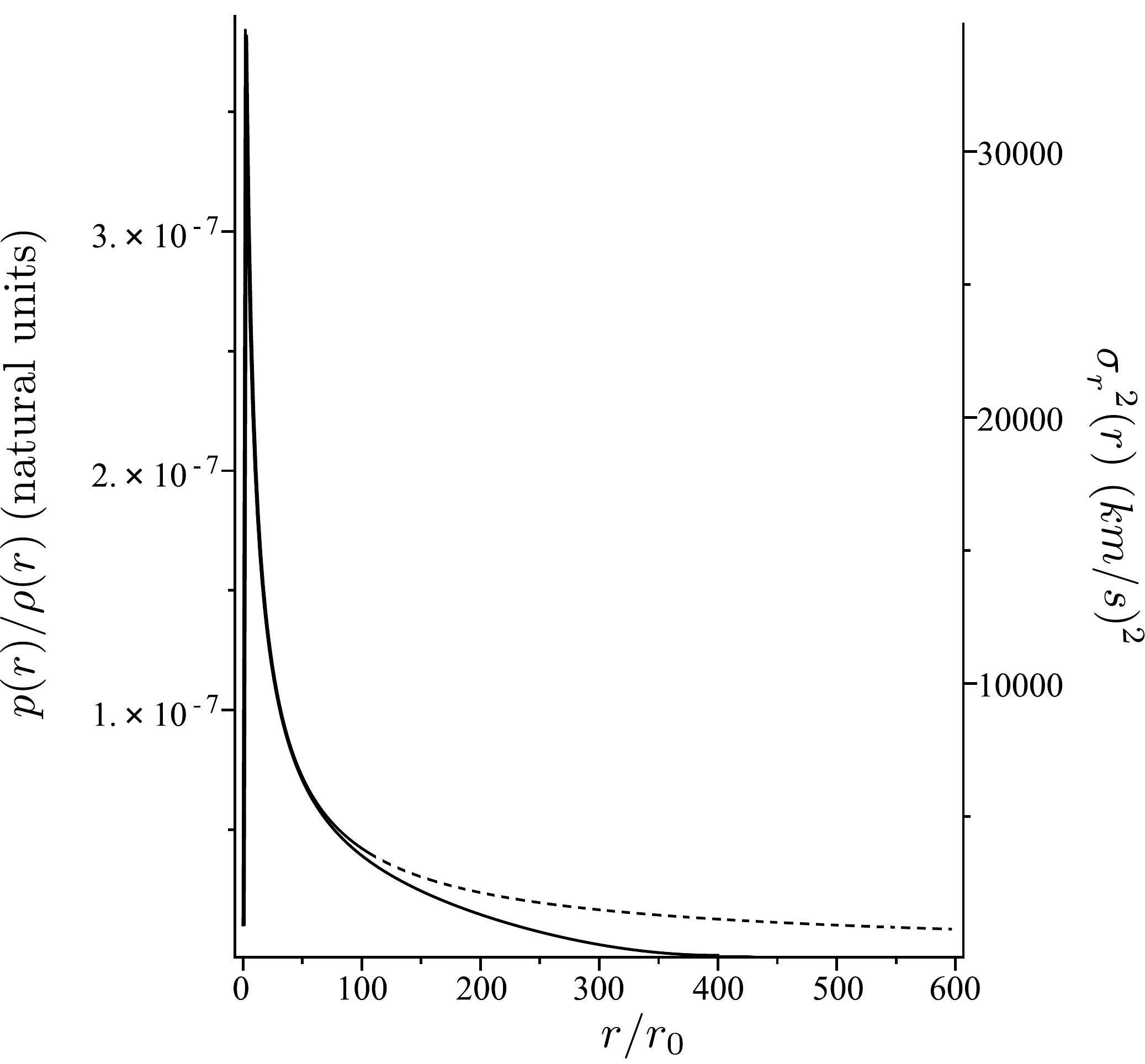}}
\end{minipage}
&
\begin{minipage}{2.5in}
\centering
\fbox{\includegraphics[height=1.6in,width=2.5in]
{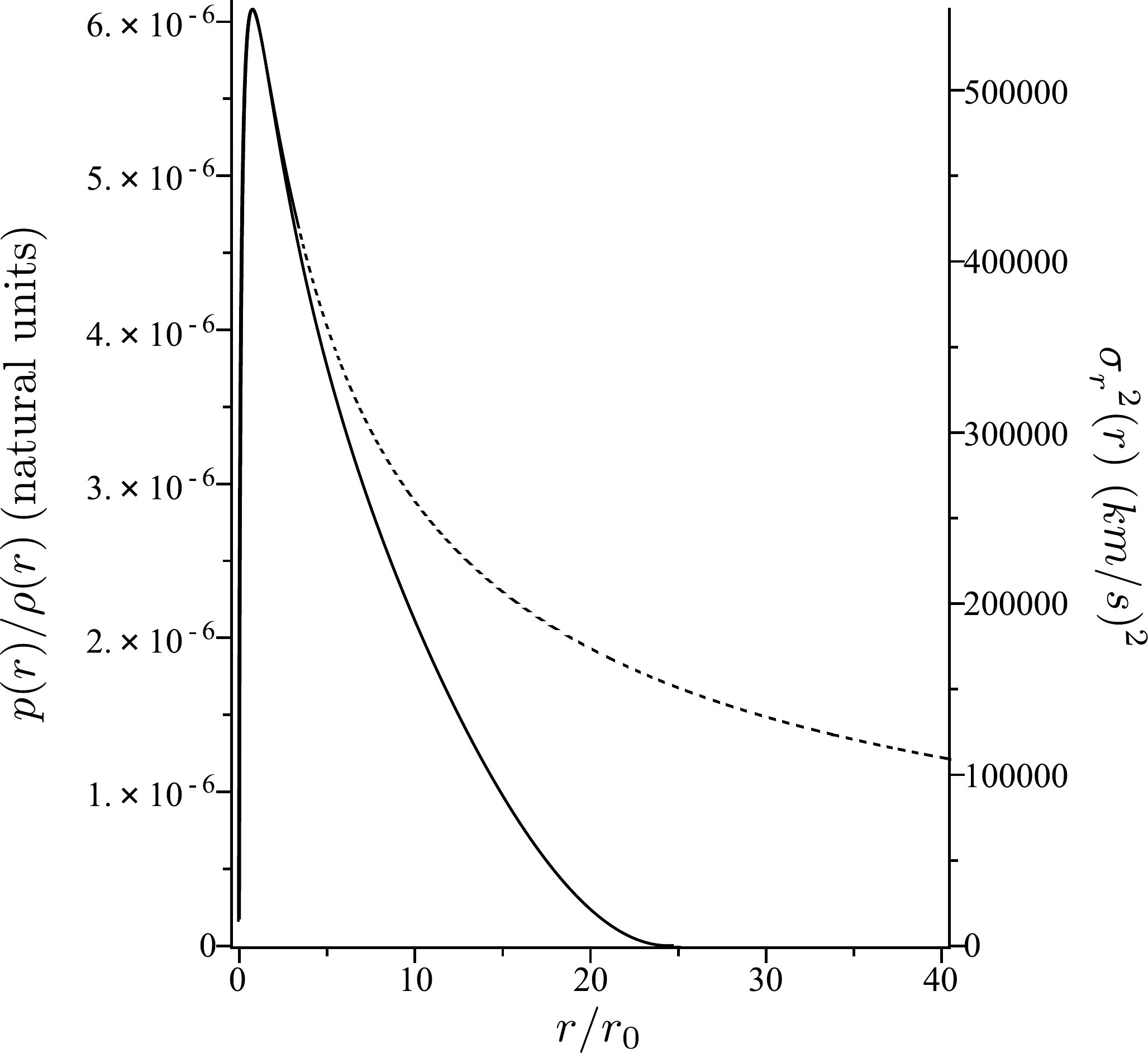}}
\end{minipage}
\\
\newline
\\
(a) Galaxy & 
(b) Cluster 
\end{tabular}
\caption{Temperature profiles $p(r)/\rho(r)$ for a galaxy and a
  cluster with NFW density profiles obtained using Newtonian theory
  assuming $\Lambda = 10^{-35}$ s$^{-2}$ (solid line) and
  $\Lambda=0$ (dotted line). Note that the solid lines actually rise again beyond their minima, but since these results are unphysical we have omitted them in the plots. \label{fig:pressures}}
\end{figure*}

Turning then to the broader features of the temperature profiles, for
$\Lambda=0$, the ratio $p(r)/\rho(r)$ is plotted in
Fig.~\ref{fig:pressures} for a galaxy and a cluster (dotted lines),
and exhibits an initial increase away from the centre, to the peak
just described, and then a gradual tailing off as $r\to \infty$. This
should be compared with the corresponding temperature profiles shown
in Fig. \ref{fig:pressures} for $\Lambda\neq 0$ (solid
lines). We see that the introduction of a non-zero $\Lambda$ has a
similar effect on the temperature profile $p(r)/\rho(r)$ as it had on
the circular velocity profile $v(r)$, namely it introduces a definite
cut-off radius at which the temperature (and pressure) vanish. Since
both $p(r)$ and $dp(r)/dr$ vanish at $r=r_F$, the temperature profile
$p(r)/\rho(r)$ also vanishes there.

Finally, rearranging equation (\ref{eq:sigmarel}) and
employing the equivalence of the ratios given by equation
(\ref{eq:prhoequivalence}), it is found that $p(r)/\rho(r)$ is also
related to the velocity dispersion of the dark matter particles through
\begin{equation}
\sigma_r^{2}(r)=c^2\left[\frac{p(r)}{\rho(r)}\right]_{\rm{nat}}=\frac{2k_{\rm
B}T(r)}{m_{\rm{p}}},\label{eq:veldispratio}
\end{equation}
where the right-most-side is obtained using (\ref{eq:gastemp}).  Since
this virial relation is independent of the mass of the dark matter
particles, it is expected to be applicable to all constituents of the
overall object which behave like collisionless particles in the
gravitational potential. Using it to calculate the peak velocity
dispersion of galaxies within a cluster, using the peak temperature
given by (\ref{eq:peaktemp}) we
obtain $\sigma_{\rm{max}}\approx 740$ km/s, which is of the same order of
magnitude as the averages found by \citet{fleenor}.  The overall scale
for $\sigma_r^{2}(r)$ through a whole galaxy/cluster is shown on a
second vertical axis in Fig.~\ref{fig:pressures}.
%REMOVE THIS SPACE

Even from this simple Newtonian analysis, in which the
expanding universe is modelled rather crudely as a de Sitter
background, comparing Tables~\ref{table:typicalnos} and
\ref{table:factors} we see that $r_F \sim 2r_T$ for clusters and $r_F\sim 4 r_T$ for galaxies, which is very encouraging.  In addition, for galaxies the values of $r_F$ are also in broad agreement with the `cut-off' radius
at which the radial velocity field vanishes in the neighbourhood of
M81, as found observationally by \citet{peirani2, peirani1} (see
Fig.~\ref{fig:peirpech}).  Given the success of
our basic Newtonian approach, in the next section we investigate the
effects of general-relativistic corrections on our analysis.

\section{General-relativistic de Sitter analysis}\label{sec:DSGR}
 
We now extend our analysis to include general-relativistic effects to
obtain a more accurate estimate of the radius $r_F$ for galaxies and
clusters and also to study further the pressure profile in such objects. We
again assume a de Sitter background universe.

Einstein's field equations with the inclusion of a non-zero
cosmological constant are 
\begin{equation}
R_{\mu\nu}-\tfrac{1}{2}g_{\mu\nu}R=\kappa T_{\mu\nu}+\Lambda
g_{\mu\nu},\label{eq:einstein}
\end{equation}
where $\kappa=8\pi$, $T_{\mu\nu}$ is the matter energy-momentum
tensor, $R_{\mu\nu} \equiv {R^{\rho}}_{\mu\rho\nu}$ and $R$ denote the
Ricci tensor and scalar respectively, and we adopt the metric
signature $(+, -, -, -)$.  We use two approaches to solve these
equations and find an expression for the speed $v(r)$ of a test
particle in a circular orbit of coordinate radius $r$ about a massive
object, as measured by a stationary observer at that radius. First, we
linearise the field equations, which necessitates assuming a
pressureless fluid within the massive object, but yields an analytical
form for $v(r)$. Second, we consider the full non-linear field
equations and again include the effect of pressure; this requires a
numerical solution to obtain $v(r)$ and the fluid pressure profile
$p(r)$ within the massive object. In both cases, we restrict the
central object to be spherically symmetric and again denote
$M(r)=\int_{0}^{r}4\pi \bar{r}^{2}\rho(\bar{r})d\bar{r}$.

\subsection{Linearised field equations}

The linearised field equations, with their attendant Lorenz gauge
constraint, are often written in the form
\begin{eqnarray}
\square^2\bar{h}^{\mu\nu} & = & -2(\kappa T^{\mu\nu}+\Lambda\eta^{\mu\nu}),\nonumber\\
\partial_{\mu}\bar{h}^{\mu\nu} & = & 0, \label{eq:LEFE}
\end{eqnarray}
where $g_{\mu\nu}=\eta_{\mu\nu}+h_{\mu\nu}$, $\eta_{\mu\nu}$ is the
Minkowski metric of flat spacetime and $h_{\mu\nu}$ is a small
perturbation \citep{GRbook}.  Also the d'Alembertian operator is defined by $\square^2\equiv
\partial_{\sigma}\partial^{\sigma}$, $\bar{h}_{\mu\nu}\equiv
h_{\mu\nu}-\frac{1}{2}\eta_{\mu\nu}h$ and $h\equiv h^{\sigma}_{\sigma}$.  One must realise, however, that this form represents the field equations specifically in Cartesian coordinates $(x,y,z)$, whereas we wish to work in spherical polar coordinates $(r,\theta,\phi)$.  In general one may interpret
the linearised field equations as describing a rank-2, symmetric
`gravitational field' $h_{\mu\nu}$ in Minkowski spacetime, so to convert them (and the Lorentz gauge condition) to curvilinear coordinates in the same
spacetime one simply makes the replacements
$\eta_{\mu\nu} \to \gamma_{\mu\nu}$ and $\partial_\mu \to \nabla_\mu$
throughout \citep{gravitation}.  Here $\gamma_{\mu\nu}$ is the Minkowski metric in
curvilinear coordinates and $\nabla_\mu$ is the covariant derivative
defined in terms of $\gamma_{\mu\nu}$, namely
\begin{align}
\nabla_{\mu}\bar{h}^{\mu\nu}&=\partial_{\mu}\bar{h}^{\mu\nu}+\Gamma^{\mu}_{\ \rho\mu}\bar{h}^{\rho\nu}+\Gamma^{\nu}_{\ \rho\mu}\bar{h}^{\mu\rho},\nonumber\\
\Gamma^{\nu}_{\ \rho\mu}&=\tfrac{1}{2}\gamma^{\nu\sigma}(\partial_{\rho}\gamma_{\sigma\mu}+\partial_{\mu}\gamma_{\rho\sigma}-\partial_{\sigma}\gamma_{\rho\mu}).
\end{align}
Working in spherical polar coordinates we specifically use $\gamma_{\mu\nu} = \mbox{diag}(1,-1,-r^2,-r^2\sin^2\theta)$. 

\subsubsection{Tetrad method}

We now obtain a solution to the linearised field equations
(\ref{eq:LEFE}) for a static, spherically-symmetric, pressureless mass
distribution $M(r)$ using a tetrad-based approach.  This method will also prove useful in solving the
full non-linear field equations and provides consistency with our
approach in NLH1.  Note that a full explanation behind this method is given in NLH1, so we skip straight to the tetrad definitions here.  

For a spherically-symmetric system in the linearised case, we may
assume a form for the tetrads obtained simply by perturbing those for
the flat Minkowski metric in spherical polar coordinates by unknown
functions $a(r)$, $b(r)$ and $c(r)$:
\begin{align}
e_{0}^{\ 0}&=1+a(r),&
e^0_{\ 0}&=1/(1+a(r)),\nonumber\\
e_1^{\ 1}&=1+b(r),&
e_{\ 1}^{1}&=1/(1+b(r)),\nonumber\\
e_2^{\ 2}&=(1+c(r))/r,&
e_{\ 2}^{2}&=r/(1+c(r)),\nonumber\\
e_3^{\ 3}&=(1+c(r))/(r\sin\theta),&
e_{\ 3}^{3}&=r\sin\theta/(1+c(r)).\label{eq:tetradslinear}
\end{align}
For a test particle with general four-velocity $\bmath{u}$, the
components in the coordinate basis
$u^\mu=[\dot{t},\dot{r},\dot{\theta},\dot{\phi}]$ are related to those
in the tetrad frame (denoted by hats) by $u^\mu={e_i}^{\mu}\hat{u}^i$, where dots denote
derivatives with respect to the particle's proper time $\tau$. Thus,
we see that for a particle (or observer) at rest in the tetrad frame,
so that $\hat{u}^i = [1,0,0,0]$, then $\dot{t}=1+a(r)$ and
$\dot{r}=\dot{\theta}=\dot{\phi}=0$. Therefore our tetrad frame
defines the local laboratory of an observer at fixed spatial
coordinates. Moreover, the three spacelike tetrad unit vectors
$\hat{\bmath{e}}_i$ $(i=1,2,3)$ lie in the same directions as the
spatial coordinate basis vectors $\bmath{e}_\mu$ $(\mu=1,2,3)$.

The line-element corresponding to the tetrads (\ref{eq:tetradslinear})
is given by
\begin{equation}
ds^2=\left[\frac{1}{1+a(r)}\right]^2dt^2
-\left[\frac{1}{1+b(r)}\right]^2dr^2 -
\left[\frac{r}{1+c(r)}\right]^2 d\Omega^2.\label{eq:linearg}
\end{equation}
where $d\Omega^2=d\theta^2+\sin^2\theta\,d\phi^2$. To first-order in
the perturbation functions, we thus have
\begin{equation}
h^{\mu\nu}=\mbox{diag}[-2a(r), \,2b(r), \,2c(r)/r^2, \,2c(r)/(r^2\sin^2\theta)].
\end{equation}
Substituting this form into the field equations and Lorenz gauge
constraint (\ref{eq:LEFE}) (appropriately modified to spherical polar
coordinates) and taking the energy-momentum tensor to be that of a
perfect fluid with zero pressure, yields the coupled differential
equations (momentarily dropping the explicit dependencies on $r$ for
brevity)
\begin{align}
a_2-b_2-2c_2+\frac{2}{r}(a_1-b_1-2c_1) + 16 \pi\rho+ 2\Lambda & =0,
\nonumber \\ a_2-b_2+2c_2+\frac{2}{r}(a_1-b_1+2c_1)
+\frac{8}{r^2}(b-c)- 2\Lambda &= 0, \nonumber \\
a_2+b_2+\frac{2}{r}(a_1+b_1)-\frac{4}{r^2}(b+c)-2\Lambda &=0,
\end{align}
where $a_1 \equiv da(r)/dr$, $a_2 \equiv d^2a(r)/dr^2$, and similar
for the derivatives of $b(r)$ and $c(r)$, and the Lorenz gauge
condition becomes
\begin{equation}
-a_1+b_1-2c_1+\frac{4}{r}(b+c)=0.
\end{equation}
Solving these equations for the three unknown functions $a(r)$, $b(r)$
and $c(r)$ gives
\begin{align}
a(r) &= -\int\frac{I_1(r)+C_1}{r^2}dr+C_2,\nonumber\\ 
b(r) &= I_3(r)+ C_4,\nonumber\\ 
c(r) &=\frac{I_2(r)}{2r^3}+\frac{C_3}{2r^3}-\frac{I_1(r)}{2r}-\frac{C_1}{2r}
+I_3(r)-\frac{\Lambda r^2}{2}+ C_4,
\label{eq:linearfcns}
\end{align}
where we have defined the integral functions
\begin{align}
I_1(r) &=\int [4\pi \rho(r) - \Lambda]r^2\,dr \nonumber \\
I_2(r) &=\int [2r I_1(r) + 2rC_1+4\pi \rho(r) r^4 + 3\Lambda r^4]\,dr 
\nonumber \\
I_3(r) &=\int\frac{I_2(r)+C_3}{r^4} \,dr,
\end{align}
and the $C_i$ ($i=1,2,3,4$) are constants of integration.

\subsubsection{Circular motion in the equatorial plane}

We now derive an expression in terms of the tetrads
(\ref{eq:tetradslinear}) for the speed $v(r)$ of a test particle in a
circular orbit of coordinate radius $r$ about the central mass, as
measured by an observer at rest at that radius (i.e. the observer
defined by our tetrad frame).

In particular, we consider a test particle moving 
along a circular path ($\dot{r}=0$) in the equatorial plane
($\theta=\pi/2$).  For such a particle, the components of the
four-velocity in the tetrad frame may be written
\begin{equation}
\hat{u}^i=[\cosh\psi(\tau),0,0,\sinh\psi(\tau)],\label{eq:fourvelLLF}
\end{equation}
where $\psi(\tau)$ is the particle's rapidity in this frame, which
may, in general, be a function of the particle's proper time $\tau$.
The particle's 3-speed, or circular velocity, $v$ in this frame is
simply $v = \hat{u}^3/\hat{u}^0 = \tanh\psi$.

One may obtain an expression for $v$ in terms of the 
tetrads (\ref{eq:tetradslinear}) most simply by using the geodesic equations
in the tetrad frame directly, which read (see NLH1)
\begin{equation}
\dot{\hat{u}}^i+\omega^i_{\ jk}\hat{u}^j\hat{u}^k = 0,
\label{eq:tetradgeo}
\end{equation}
where the spin-connection ${\omega^i}_{jk}$ is given by
\begin{align}
\omega_{ijk}&=\tfrac{1}{2}(c_{ijk}+c_{jki}-c_{kij}),\nonumber\\
{c^{k}}_{ij}&={e_i}^{\mu}{e_j}^{\nu}(\partial_{\mu}{e^k}_{\nu}
-\partial_{\nu}{e^k}_{\mu}).\label{eq:cdefn}
\end{align}
Using the tetrads (\ref{eq:tetradslinear}), the non-zero elements of the
spin-connection are ${\omega^1}_{00} = {\omega^0}_{10}$ and 
${\omega^1}_{22} = {\omega^1}_{33} = -{\omega^2}_{12} = -{\omega^3}_{13}$, where
\begin{eqnarray}
{\omega^1}_{00} & = & -\frac{1+b(r)}{1+a(r)}\nd{a(r)}{r},
\nonumber\\
{\omega^1}_{22} & = &
\frac{1+b(r)}{r(1+c(r))}\left(r\nd{c(r)}{r}-c(r)-1\right).
\label{eq:spinconns}
\end{eqnarray}
Substituting these expressions into (\ref{eq:tetradgeo}) and using
(\ref{eq:fourvelLLF}), one finds that the geodesic equation for $i=2$
is satisfied identically and the equations for $i=0$, 3 both reduce to
$\dot{\psi} = 0$, which indicates that, as expected, the particle has
a constant rapidity, and hence constant 3-speed, in the tetrad
frame. Finally, the geodesic equation for $i=1$ gives $\tanh^2\psi =
-{\omega^1}_{00}/{\omega^1}_{33}$, from which we immediately obtain
\begin{equation}
v(r)=\left[\frac{r(1+c(r))\frac{da(r)}{dr}}{(1+a(r))\left(\frac{dc(r)}{dr}-c(r)-1\right)}\right]^{\frac{1}{2}}.\label{eq:tanh}
\end{equation}

%It is algebraically simpler and more instructive, however, first to
%relate the components of the four-velocity in the tetrad frame to
%those in the coordinate frame by
%$u^\mu=e_i^{\ \mu}\hat{u}^i=[\dot{t},\dot{r},\dot{\theta},\dot{\phi}]$. 
%In particular, using the tetrad definitions (\ref{eq:tetradslinear}), one finds
%\begin{equation}
%v = \frac{\hat{u}^3}{\hat{u}^0} = \frac{r(1+a(r))}{1+c(r)}
%\frac{\dot{\phi}}{\dot{t}} = \frac{r(1+a(r))}{1+c(r)} \nd{\phi}{t}
%\label{eq:vdphidt}
%\end{equation}
%One can then easily obtain an expression for the coordinate angular velocity
%$d\phi/dt$ from the geodesic equations in the coordinate frame, which are
%most conveniently expressed as \citep{GRbook}
%\begin{equation}
%\dot{u}_{\mu}=\tfrac{1}{2}(\partial_{\mu}g_{\nu\sigma})u^{\nu}u^{\sigma}.
%\end{equation}
%For circular motion in the equatorial plane, one has
%$d\phi/dt=u^3/u^0$ and $u^1=u^2=0$. Hence
%$u_1=g_{11}u^1=0$ and $\dot{u}_1=0$, and one quickly finds
%\begin{equation}
%\left(\nd{\phi}{t}\right)^2 =
%-\frac{\partial_1g_{00}}{\partial_1g_{33}} =
%\frac{\frac{da(r)}{dr}(1+c(r))^3}{r(1+a(r))^3\left(r\frac{dc(r)}{dr}-1-c(r)\rig%ht)}.\label{eq:dphidt}
%\end{equation}
%Thus, combining (\ref{eq:vdphidt}) and (\ref{eq:dphidt}), the
%expression for the speed $v(r)$ of the particle, as measured by our
%stationary observer, is 
%\begin{equation}
%v(r)=\left[\frac{r(1+c(r))\frac{da(r)}{dr}}{(1+a(r))\left(\frac{dc(r)}{dr}-c(r)%-1\right)}\right]^{\frac{1}{2}}.\label{eq:tanh}
%\end{equation}

\subsubsection{Results for example radial density profiles}

Since the circular velocity is expected to behave sensibly everywhere,
the functions $a(r)$, $b(r)$ and $c(r)$ should not diverge at the
origin. By considering how their Taylor expansions about $r=0$ may
satisfy this requirement, it is not difficult to solve for the
constants $C_i$ ($i=1,2,3,4$) for a specific radial density profile
$\rho(r)$. Expressions for the constants for the exponential profile
(\ref{eq:expprofile}) and the NFW profile (\ref{eq:nfwprofile}) are
given in Table~\ref{table:densprof}.
\begin{table}
\begin{center}
\begin{tabular}{@{}ccc@{}}\hline
& Exponential $\rho(r)$ & NFW $\rho(r)$ \\ \hline
$C_1$ & $8\pi \rho_0 r_0^3$ & $-4\pi\rho_0r_0^3(\ln r_0+1)$ \\
$C_2$ &  $-4\pi\rho_0 r_0^2$ & $\tfrac{1}{6}\Lambda r_0^2-4\pi\rho_0r_0^2$ \\
$C_3$ & 0 & 0 \\
$C_4$ & $4\pi\rho_0 r_0^2$ & $\tfrac{7}{30}\Lambda r_0^2 + 4\pi\rho_0 r_0^2$ \\
\hline
\end{tabular}
\caption{Values of the constants in the functions $a(r)$, $b(r)$ and $c(r)$ in (\ref{eq:linearfcns}) for each considered density profile.}\label{table:densprof}
\end{center}
\end{table}
The resulting circular velocity $v(r)$ profiles are practically
identical to those plotted in Fig.~\ref{fig:newtonian}, and so are not
presented separately.  This supports our previous results and
indicates that linear general-relativistic corrections to the
(circular) orbits of test particles are unimportant for massive
objects with densities typical of galaxies and clusters, as might be
expected. Nonetheless, the analytical results we have derived may
prove useful when general-relativistic effects for astronomical
extended density distributions become measurable.

\subsection{Full non-linear field equations}

We now consider the full non-linear Einstein equations and model the
massive object as a perfect fluid with some (non-zero) radial pressure
profile $p(r)$, as we did for the Newtonian case. To solve the field
equations, we again use the tetrad-based method described in NLH1 and
employed in the previous section. Fortunately, our task is somewhat
simpler for the full field equations than for the linearised
equations, since we already considered general, time-dependent,
spherically-symmetric systems in this case in NLH1 (see also \citet{GGTGA}), which we now summarise.  

We have considered a general metric element in terms of three unknown functions $g_1(r,t)$, $g_2(r,t)$ and $f_1(r,t)$,
\begin{equation}
ds^2=\left(\frac{g_1^{\ 2}-g_2^{\ 2}}{f_1^{\ 2}g_1^{\ 2}}\right)dt^2+\frac{2g_2}{f_1g_1^{\ 2}}\,dr\,dt-\frac{1}{g_1^{\ 2}}\,dr^2-r^2d\Omega^2.\label{eq:generalmetric}
\end{equation}
We have then shown that, assuming the matter to be a perfect fluid with density $\rho(r,t)$ and $p(r,t)$, Einstein's field equations and the Bianchi identities yield the following relationships between the unknown quantities:
\begin{align}
L_rf_1&=-Gf_1\Rightarrow f_1=\exp\left\{-\int^r\frac{G}{g_1}dr\right\},\nonumber\\
L_r g_1&=F g_2+\frac{M}{r^2}-\tfrac{1}{3}\Lambda r-4\pi r
\rho,\nonumber\\
L_t g_2&=G g_1-\frac{M}{r^2}+\tfrac{1}{3}\Lambda r-4\pi r p,\nonumber\\
L_t M&=-4\pi g_2 r^2p,\nonumber\\
L_t\rho&=-\left(\frac{2g_2}{r}+F\right)(\rho+p),\nonumber\\
L_rM&=4\pi g_1 r^2\rho,\nonumber\\
L_r p&=-G(\rho+p).\label{eq:alleqns}
\end{align}
Here we have defined two linear
differential operators
\begin{align}
L_t&\equiv f_1\partial_t+g_2\partial_r,\nonumber\\
L_r&\equiv g_1\partial_r,
\end{align}
and the function $F(r,t)$, the radial acceleration, $G(r,t)$, and the intrinsic mass (or energy) interior to $r$, $M(r,t)$, by
\begin{align}
L_t g_1&\equiv G g_2,\nonumber\\
L_r g_2 &\equiv F g_1,\nonumber\\
M&\equiv \tfrac{1}{2}r\left(g_2^{\ 2}-g_1^{\ 2}+1-\tfrac{1}{3}\Lambda r^2\right),\label{eq:quantities}
\end{align}
where $\Lambda$ is the cosmological constant.

In this work, rather than considering a point mass as in NLH1, we are considering a static matter distribution (but with
a non-zero cosmological constant).  Thus all functions depend only on
the radial coordinate $r$, and the operator $L_t$ reduces simply to
$L_t = g_2(r)\partial_r$. For a given radial density profile
$\rho(r)$, one has $M(r) = \int_0^r 4\pi \bar{r}^2
\rho(\bar{r})\,d\bar{r}$, as previously, and our goal is to determine
the unknown functions $f_1(r)$, $g_1(r)$, $g_2(r)$, $F(r)$, $G(r)$ and
$p(r)$. 
%It is clear from the $L_r f_1$ equation in (\ref{eq:alleqns}),
%however, that one may immediately obtain $f_1(r)$ once $g_1(r)$ and $G(r)$
%have been determined.

Substituting the above expression for $M(r)$ into the $L_tM$ equation
in (\ref{eq:alleqns}), one finds $g_2(r)\rho(r) = -g_2(r)p(r)$. Since
the density and pressure must be positive in a realistic object, one
immediately concludes that $g_2(r)=0$. It is worth noting at this
point that only the `diagonal' tetrads ${e_i}^i$ (no sum
on $i$) defined in NLH1 and their inverses are non-zero. Thus,
the general form of the line-element has reduced to the
simple form
\begin{equation}
ds^2=\frac{1}{f_1^{\ 2}(r)}dt^2-\frac{1}{g_1^{\ 2}(r)}dr^2-r^2d\Omega^2.
\label{eq:ginterior}
\end{equation}
Moreover, for a particle (or observer) at rest in the tetrad frame,
so that its four-velocity components are $[\hat{u}^i] = [1,0,0,0]$, we
thus have $\dot{t}=f_1(r)$ and $\dot{r}=\dot{\theta}=\dot{\phi}=0$.
Therefore, once again, our tetrad frame defines the local laboratory
of an observer at fixed spatial coordinates and the three spacelike
tetrad unit vectors $\hat{\bmath{e}}_i$ $(i=1,2,3)$ lie in the same
directions as the spatial coordinate basis vectors
$\bmath{e}_\mu$ $(\mu=1,2,3)$.

Since $g_2(r)=0$, the $M$ and $L_rg_2$ equations in
(\ref{eq:quantities}), and the $L_tg_2$ equation in (\ref{eq:alleqns})
yield, respectively,
\begin{eqnarray}
g_{1}(r) & = & \sqrt{1-\frac{2M(r)}{r}-\tfrac{1}{3}\Lambda r^2}, \nonumber \\
F(r) & = & 0 \nonumber \\
G(r) & = & \frac{1}{g_{1}(r)}\left[\frac{M(r)}{r^2}-\tfrac{1}{3}\Lambda r 
+4\pi rp(r)\right]
\label{eq:fullsols}
\end{eqnarray}
Moreover, once $g_1(r)$ and $G(r)$ have been determined, one may
immediately obtain $f_1(r)$ from the $L_r f_1$ equation in
(\ref{eq:alleqns}). Thus the solution is fully specified by the three
non-zero functions $g_1(r)$, $G(r)$ and $p(r)$.

We already have an explicit expression for $g_1(r)$ in terms of known
quantities, and an explicit expression for $G(r)$ that depends on the
unknown function $p(r)$. The pressure profile must satisfy the $L_rp$ equation
in (\ref{eq:alleqns}), which reads
\begin{equation}
\frac{dp(r)}{dr}={-\frac{G(r)}{g_{1}(r)}[\rho(r)+p(r)]}.
\label{eq:grdpdr}
\end{equation}
Combining this equation with the $G(r)$ equation in
(\ref{eq:fullsols}), one obtains the Oppenheimer--Volkov equation with
a cosmological constant \citep{winter},
\begin{equation}
\frac{dp(r)}{dr}=-\frac{(\rho(r)+p(r))(M(r)+4\pi r^3p(r)-\frac{1}{3}\Lambda r^3)}{r\left(r-2M(r)-\frac{1}{3}\Lambda r^3\right)},\label{eq:OpVol}
\end{equation}
which directly relates the pressure and density profiles of a static,
spherically-symmetric perfect fluid distribution.  The
Oppenheimer--Volkov equation may be considered as the relativistic
generalisation of the Newtonian equation of hydrostatic equilibrium
(\ref{eq:OVNewtLambda}), to which it reduces in the appropriate
limits. For a given
$\rho(r)$ and suitable boundary condition on the pressure, we may 
solve (\ref{eq:OpVol}) to obtain $p(r)$, and hence $G(r)$.

It is worth noting that in the special case of a point mass, such that
$M(r)=m=\text{constant}$, $\rho(r)=\delta(r)$ and $p(r)=0$, one quickly finds
\begin{align}
g_1(r)&=\sqrt{1-\frac{2m}{r}-\frac{1}{3}\Lambda r^2},\nonumber\\
G(r)&=\frac{\frac{m}{r^2}-\frac{1}{3}\Lambda r}{\sqrt{1-\frac{2m}{r}-\frac{1}{3}\Lambda r^2}},\nonumber\\
f_1(r)&=\frac{1}{\sqrt{1-\frac{2m}{r}-\frac{1}{3}\Lambda r^2}}.
\end{align}
The corresponding line-element (\ref{eq:ginterior}) in this case reads
\begin{align}
ds^2&=\left(1-\frac{2m}{r}-\tfrac{1}{3}\Lambda r^2\right)dt^2 \nonumber \\
& \hspace{1.8cm}
-\left(1-\frac{2m}{r}-\tfrac{1}{3}\Lambda r^2\right)^{-1}dr^2 -r^2d\Omega^2,
\label{eq:SDS}
\end{align}
which is the Schwarzschild--de Sitter metric, as expected.

\subsubsection{Boundary condition on the pressure profile}\label{sec:boundaries}

The boundary condition on $p(r)$ required to solve the
Oppenheimer--Volkov equation (\ref{eq:OpVol}) may be calculated to a
high degree of accuracy by considering a series expansion of $p(r)$
about $r=0$. For a given density profile $\rho(r)$, we can deduce a
suitable form for this expansion by considering the
Oppenheimer--Volkov equation in its Newtonian limit with $\Lambda$ set
to zero, namely
\begin{equation}
\frac{dp(r)}{dr}=-\frac{\rho(r)M(r)}{r^2},\label{eq:OVNewtnoLambda}
\end{equation}
which is the Newtonian equation of hydrostatic equilibrium
(\ref{eq:OVNewtLambda}) with $\Lambda=0$.

For the exponential density profile (\ref{eq:expprofile}), $\rho(r) \to
\mbox{constant}$ as $r\to 0$, and hence $M(r) \propto r^3$. From
(\ref{eq:OVNewtnoLambda}), we thus see that $p(r) \propto r^2$ as $r
\to 0$.  Thus, one may adopt a simple Taylor expansion
$p(r)=\sum_{n=0}^\infty p_n r^n$ about $r=0$. For the NFW density
profile (\ref{eq:nfwprofile}), however, more care is required in
determining the appropriate expansion, since $\rho(r) \propto 1/r$ as
$r \to 0$. Therefore, in this limit, $M(r) \propto r^2$ and from
(\ref{eq:OVNewtnoLambda}) we have $p(r) \propto \ln r$. This
logarithmic divergence at the origin must now be taken into account by
instead using a power series of the form
\begin{equation}
p(r) = \sum_{n=0}^\infty \left[p_n +
  \tilde{p}_n\ln\left(\frac{r}{r_0}\right)\right]r^n.
\end{equation}
For each density profile $\rho(r)$, the coefficients in the expansion
for $p(r)$ can be found by substituting the corresponding series into
the Oppenheimer--Volkov equation (\ref{eq:OpVol}). These coefficients
are then used to set the boundary condition on the pressure at
$r=10^{-3}r_0$, from which point (\ref{eq:OpVol}) is numerically
integrated outwards to obtain $p(r)$. As in the Newtonian case, we
impose the boundary condition that the pressure itself vanishes where
$dp(r)/dr=0$.

\subsubsection{Circular motion in the equatorial plane}

We again consider a test particle moving along a circular path
($\dot{r}=0$) in the equatorial plane ($\theta=\pi/2$), for which the
components of the four-velocity in the tetrad frame may be written as
in (\ref{eq:fourvelLLF}). Once again, one may obtain an expression for the
3-speed $v=\tanh\psi$ of the test particle 
in the tetrad frame by using the
geodesic equations (\ref{eq:tetradgeo}). In this case, the non-zero
elements of the spin-connection (\ref{eq:cdefn}) are again
${\omega^1}_{00} = {\omega^0}_{10}$ and ${\omega^1}_{22} =
{\omega^1}_{33} = -{\omega^2}_{12} = -{\omega^3}_{13}$, where
\begin{align}
{\omega^1}_{00} &=  -\frac{g_1(r)}{f_1(r)}\nd{f_1(r)}{r} = G(r),
\nonumber\\
{\omega^1}_{22} &= - \frac{g_1(r)}{r},
\label{eq:spinconns2}
\end{align}
where, for ${\omega^1}_{00}$, in the second equality we have used the
$L_r f_1$ equation in (\ref{eq:alleqns}).

We again find that the resulting geodesic equation for $i=2$ is
satisfied identically, that the equations for $i=0$, 3 both reduce to
$\dot{\psi} = 0$, and that the equation for $i=1$ gives $\tanh^2\psi =
-{\omega^1}_{00}/{\omega^1}_{33}$. Thus the speed $v(r)$ of a particle
in a circular orbit at coordinate radius $r$, as measured by a stationary 
observer at that radius is
\begin{equation}
v(r)= \sqrt{\frac{rG(r)}{g_1(r)}}.
\label{eq:circvelfullgr}
\end{equation}

\subsubsection{Results for example density profiles}

The resulting circular velocity profiles $v(r)$ and pressure profiles
$p(r)$ are again indistinguishable from those plotted in
Figs~\ref{fig:newtonian} and \ref{fig:pressures} using the Newtonian
theory, and are therefore not plotted separately.  Thus, even full
general-relativistic corrections to are unimportant for massive
objects with densities typical of galaxies and clusters \citep{hwang}.

Indeed, quite remarkably, one can show that the location of the radius
$r=r_F$ at which the circular velocity $v(r)$ and the pressure $p(r)$
vanish is {\em identical} in the Newtonian and full
general-relativistic cases, even for large densities and pressures at which
general-relativistic effects become important. From
(\ref{eq:circvelfullgr}), we see that the radius $r=r_F$ is also the
point at which the radial acceleration $G(r)$ vanishes, changing its
direction from inwards for $r < r_F$ to outwards for $r > r_F$.  From
(\ref{eq:grdpdr}), this coincides with the radius at which $dp(r)/dr$
vanishes, which is also where $p(r)$ vanishes according to our
boundary condition. Thus, evaluating the $G(r)$-equation in
(\ref{eq:fullsols}) at $r=r_F$, one finds that
\begin{equation}
\frac{M(r_F)}{r_F^2} - \tfrac{1}{3}\Lambda r_F = 0, 
\end{equation}
which is precisely the Newtonian condition defining $r_F$.

Although $r_F$ is a natural physical radius separating bound and
unbound material in the galaxy or cluster, it is unclear whether
$r=r_F$ can be interpreted as the {\itshape{maximum size}} of a
massive object, since this notion depends more on the physical
stability of particle orbits, rather than merely the point where the
radial force on the particle vanishes.

\section{General-relativistic analysis for a central point mass}
\label{sec:expansion}

We now address the question of orbit stability and also consider a
more realistic, time-dependent model for the background cosmological
expansion, albeit at the price of modelling the central massive object
as a point mass.

In NLH1, we derived the metric for a point mass $m$ embedded in an
expanding cosmological background, for spatially-flat and
spatially-curved models. In the spatially-flat case, which is a
reasonable description of our universe, and using non-comoving
(`physical') coordinates, the metric reads
\begin{align}
ds^2=&\left[1-\frac{2m}{r}-r^2H^2(t)\right]\,dt^2+2rH(t)
\left(1-\frac{2m}{r}\right)^{-\frac{1}{2}}dr\,dt\nonumber\\
& -\left(1-\frac{2m}{r}\right)^{-1}\,dr^2-r^2\,d\Omega^2.\label{eq:flat}
\end{align}
Note that this metric is only applicable outside $r> 2m$.
    Nonetheless it is appropriate to use it for our region of interest
    $r\gg m$.  As shown in NLH1, it is connected by a coordinate
transformation to the corresponding metric derived by \cite{mcvittie}.
It is worth noting that in the special case of a de Sitter background,
for which $H^2(t)=\Lambda/3$, the coordinate transformation
\citep{arakida}
\begin{equation}
dt=d\bar{t}-\frac{1}{\left(1-\frac{2m}{r}-\frac{1}{3}\Lambda r^2\right)}\sqrt{\frac{\frac{1}{3}\Lambda r^2}{1-\frac{2m}{r}}}dr.\label{eq:MVtoSDS}
\end{equation}
converts the metric (\ref{eq:flat}) into the standard
Schwarzschild--de Sitter metric (\ref{eq:SDS}) (with $t$ replaced by
$\bar{t}$), as expected.

\subsection{Radial force on a test particle}

In NLH1, we also obtained an invariant expression for the force per unit mass
required to keep a test particle at rest relative to the central mass
$m$. In the spatially-flat case, this reads
\begin{equation}
f =\frac{\frac{m}{r^2} - rH^2(t)}{\left(1-\frac{2m}{r}-r^2H^2(t)\right)^{1/2}}
+\frac{rH^2(t)(q(t)+1)\sqrt{1-\frac{2m}{r}}}
{\left(1-\frac{2m}{r}-r^2H^2(t)\right)^{3/2}}.
\label{eq:flatforcefull}
\end{equation}
In the special case of a de Sitter background, for which
$H^2(t)=\Lambda/3$ and $q(t)=-1$, this reduces to the well-known
result
\begin{equation}
f_{\text{dS}}=\frac{\frac{m}{r^2}-\frac{1}{3}\Lambda r}{\sqrt{1-\frac{2m}{r}-\frac{1}{3}\Lambda r^2}}.
\end{equation}

As discussed in NLH1, the Newtonian approximation to
(\ref{eq:flatforcefull}) in our region of interest $m \ll r \ll
1/H(t)$, and a change of sign, allows us to identify the radial force
on a unit-mass test particle as simply that given in (\ref{eq:eq1}), namely
\begin{equation}
F \approx -\frac{m}{r^2} - q(t)H^2(t)r.
\label{eq:eq1again}
\end{equation}
Thus, the force consists of the usual $1/r^2$ inwards component due to
the central (point) mass $m$ and a cosmological component proportional
to $r$ that is directed outwards (inwards) when the expansion of the
universe is accelerating (decelerating). 

The key difference between the general expression (\ref{eq:eq1again})
and the special case of a de Sitter background is that the
cosmological force component in the former is time-dependent.  Indeed,
as shown in section \ref{sec:newtlimforces}, for the standard $\Lambda$CDM concordance cosmology,
the cosmological force term reverses direction at about $z\approx
0.67$, changing from an inwards directed force at high redshift (when
the universal expansion is decelerating) to an outwards directed force
at low redshift (as the universal expansion accelerates owing to the
dominance of the dark energy component). For the standard $\Lambda$CDM
concordance cosmology, the dominance of the dark energy component will
increase as the universe continues to expand and so the expansion will
tend to the de Sitter background model considered earlier, for which
the cosmological force term is time-independent.

The time-dependence of the cosmological force term in the general case
means that the radius at which the total radial force becomes zero,
which marks the cut-off point for the existence of circular particle
orbits, is also time-dependent. For a central point mass, this is
given explicitly by
\begin{equation}
r_{F}(t)\approx\left[-\frac{m}{q(t)H^2(t)}\right]^{\frac{1}{3}},
\label{eq:accurateforce}
\end{equation}
provided the universal expansion is accelerating, so that $q(t)$ is
negative.  This also marks the cut-off between gravitationally bound and unbound material, so it may be interpreted as the point separating the `system' from the rest of the universe.  Clearly, if the expansion is decelerating (i.e. for $z >
0.67$ in the $\Lambda$CDM concordance model), then the force due to
the central mass $m$ and the cosmological force are both directed
inwards and so there is no radius at which the total force vanishes,
and hence no natural cut-off for the size of massive objects. Thus, we
see that the time-dependent cosmological force term in a general
expansion will have a profoundly different effect on the formation and
structure of massive objects as compared with the simple special case
of a time-independent de Sitter background; in the latter case, the
expression (\ref{eq:accurateforce}) reduces simply to $r_F =
(3m/\Lambda)^{1/3}$.

Nonetheless, at low redshifts, where the dark energy component is
dominant, we might expect that the values of $r_F$ obtained using
(\ref{eq:accurateforce}) will not differ significantly from those
obtained assuming a de Sitter background. The dark energy density
$\Omega_\Lambda(t) = \Lambda/[3H^2(t)]$ has the value
$\Omega_{\Lambda,0} \approx 0.7$ at the current epoch $t=t_0$ and the
deceleration parameter is currently measured at $q(t_0)\approx-0.55$
\citep{WMAP}. Thus, the `correction factor' one should apply to our
earlier results in Table~\ref{table:factors}, derived assuming a
de Sitter background, is only
$(-\Omega_{\Lambda,0}/q(t_0))^{1/3}\approx 1.08$, assuming
that the correction required in the case of an extended central mass
is approximately the same as that for a point mass. 

We may compare the present-day values $r_F(t_0)$, calculated from
(\ref{eq:accurateforce}), with the cut-off radii obtained in the numerical
analysis by \citet{peirani2,peirani1} for specific low-redshift
galaxies. This comparison is listed in Table~\ref{table:PPcomparison},
from which we see good agreement between the two approaches.
\begin{table}
\begin{center}
\begin{tabular}{@{}lcccc@{}}\hline
Galaxy & Mass ($10^{11}M_{\odot}$) & {P\&P} cut-off & $r_F(t_0)$ & $r_S(t_0)$\\ \hline
M83 & $21$ & $1.10$ & $1.47$ & $0.93$\\ 
M81 & $9.2$ & $0.75$ & $1.12$ & $0.70$\\
IC 342 & $2$ & $0.51$ & $0.67$ & $0.42$\\ 
NGC 253 & $1.3$ & $0.40$ & $0.58$ & $0.36$\\ \hline
\end{tabular}
\caption{Radial distances $r_F(t_0)$ and $r_S(t_0)$ in Mpc from the centre of each galaxy, assuming the galaxy to be a point mass, as compared to the empirical `cut-off' radius derived by Peirani \&
  Pacheco (`P\&P') at which the radial velocity field vanishes; see
  text for details.\label{table:PPcomparison}}
\end{center}
\end{table}
It should be noted, however, that the $r_F(t_0)$ values are typically
about a factor $\sim 1.4$ larger than the corresponding radial cut-offs
derived by \citet{peirani2,peirani1}. We now address this discrepancy
by considering the largest {\em stable} circular orbits possible about
the central mass.

\subsection{Largest stable orbits}

We have mentioned previously that the radius $r=r_F$ at which the
total radial force is zero, and the circular velocity $v(r)$ also
vanishes, does not necessarily correspond to the maximum possible size
of the galaxy or cluster.  Indeed many of the gravitationally-bound particles inside $r_F$ may be in unstable circular orbits.   A more physically meaningful `outer' radius, which one may interpret as the maximum size of the object,
is that corresponding to the largest {\em stable} circular orbit,
which we denote by $r_S$.

The radius $r_S$ may be determined as the minimum of the
(time-dependent) effective potential for
a test particle in orbit about the central mass $m$
\citep{GRbook}. For the metric (\ref{eq:flat}), we found in NLH1 that
the corresponding geodesic equations lead to an equation of motion for
a test particle given by
\begin{align}
\ddot{r}=&\frac{L^2}{r^3}\left(1-\frac{3m}{r}\right)-\frac{m}{r^2}+rH^2(t)-\frac{2r^2H(t)H^{\prime}(t)}{1-\frac{2m}{r}-r^2H^2(t)}\dot{r}\dot{t}\nonumber\\
&+\frac{H^{\prime}(t)r\sqrt{1-\frac{2m}{r}}}{1-\frac{2m}{r}-r^2H^2(t)}\left(1+\frac{L^2}{r^2}\right)\nonumber\\
&+\frac{rH^{\prime}(t)}{\sqrt{1-\frac{2m}{r}}\left(1-\frac{2m}{r}-r^2H^2(t)\right)}\dot{r}^2,\label{eq:fullgeo}
\end{align}
where $L$ is the specific angular momentum of the test particle about
the central mass $m$.  We recall that dots denote derivatives with
respect to the particle's proper time $\tau$, whereas primes
denote derivatives with respect to the cosmic time $t$.  

Expanding (\ref{eq:fullgeo}) in $m$ and assuming a weak gravitational
field and low velocities for the test particle ($\dot{t}\approx 1$,
$\dot{r}\approx 0$), one obtains the approximate Newtonian expression
\begin{equation}
\frac{d^2r}{dt^2} \approx -\frac{m}{r^2}-q(t)H^2(t)r
+\frac{L^2}{r^3}. \label{eq:newtgeo}
\end{equation}
This is simply the Newtonian radial force expression
(\ref{eq:eq1again}) with the inclusion of a centrifugal term. 

\subsubsection{Analytic results}

The effective potential in which the test particle moves is given by
the first integral of (\ref{eq:newtgeo}) with respect to $r$, and so
its turning points occur at the $r$-values for which $d^2r/dt^2=0$,
namely the solutions of
\begin{equation}
q(t)H^2(t)r^4+mr-L^2=0.\label{eq:potcond}
\end{equation}
The two real solutions of this quartic both include the factor
$\sqrt{256q(t)H^2(t)L^6+27m^4}\geq0$, thus placing an upper bound on
$L$ of
\begin{equation}
L\leq\left[-\frac{27m^4}{256q(t)H^2(t)}\right]^{\frac{1}{6}}.
\end{equation}
Using the maximum value of $L$, the radius of the largest stable circular
orbit is found to be
\begin{equation}
r_S(t) =\left[-\frac{m}{4q(t)H^2(t)}\right]^{\frac{1}{3}}.\label{eq:maxstable}
\end{equation}

Comparing this expression with that for $r_F(t)$ in
(\ref{eq:accurateforce}), we see that although both radii are
time-dependent, they are related by a constant factor. Thus, at any
epoch, the radius $r_S(t)$ of the largest stable circular orbit lies a
factor $4^{1/3} \approx 1.6$ inside the radius $r_F(t)$ at which the
total radial force on a test particle is zero and the circular
velocity $v(r)$ vanishes. This additional `correction factor' brings
our estimates of the maximum possible sizes of individual galaxies into
very close agreement with the cut-off radii derived by
\citet{peirani2,peirani1}, as shown in Table~\ref{table:PPcomparison}.
For a galaxy of total mass $m \approx 10^{12}M_{\odot}$, the radius
of the largest stable circular orbit at the present epoch is $r_S(t_0)
\approx 0.72$ Mpc, and for a cluster of total mass $m \approx 10^{15}M_{\odot}$ we have $r_S(t_0)\approx 7.2$ Mpc.  These values are consistent with the observed sizes of
low-redshift galaxies and clusters respectively; only $\sim2.8$ and $\sim1.4$ times larger than the typical sizes quoted in Table \ref{table:typicalnos}. We note that in the special case of a de Sitter
background, the expression (\ref{eq:maxstable}) reduces to $r_S =
(3m/(4\Lambda))^{1/3}$.

\begin{figure*}
\centering
\begin{tabular}{cc}
\begin{minipage}{2.5in}
\centering
\fbox{\includegraphics[height=1.6in,width=2.2in]
{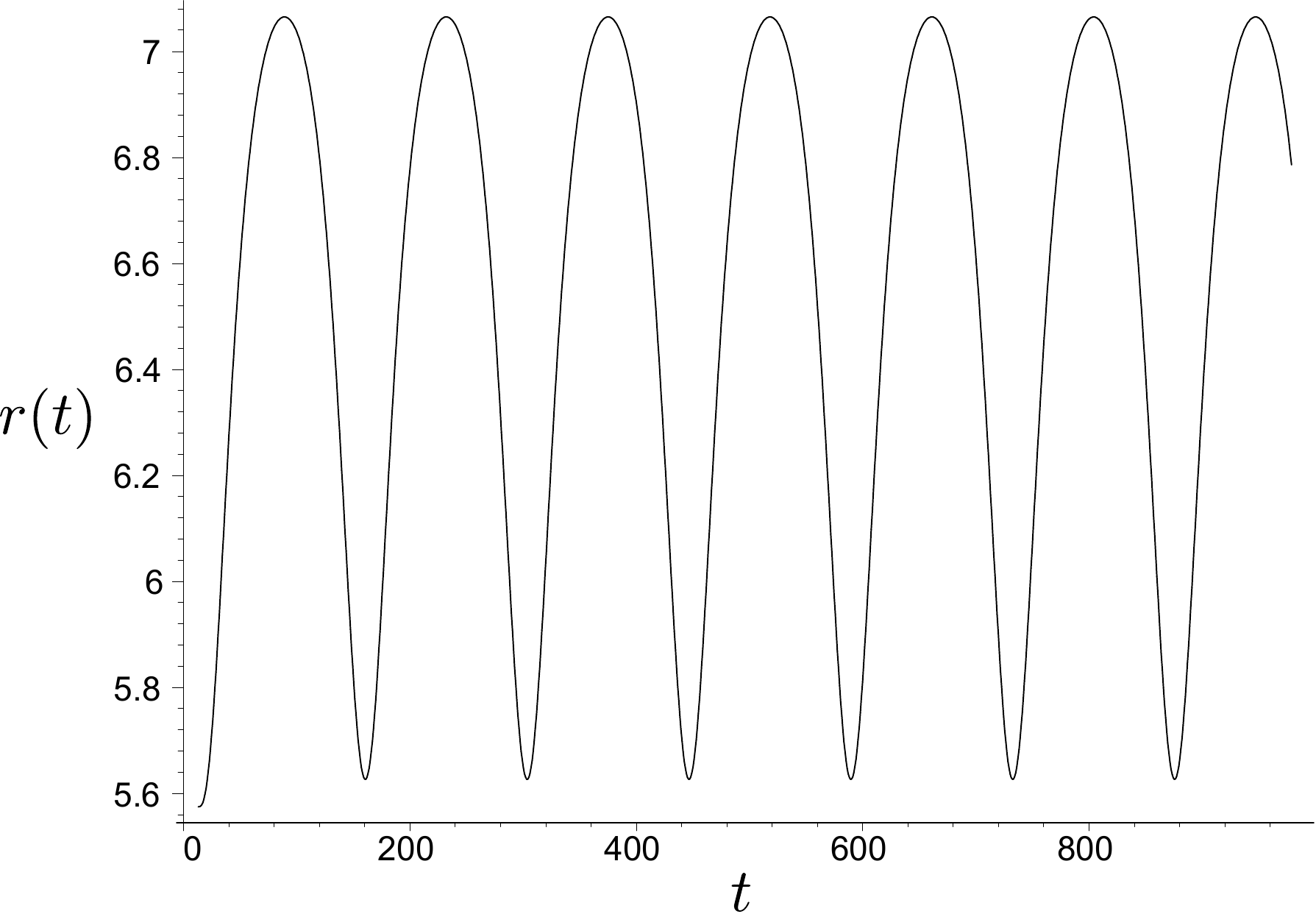}}
\end{minipage}
&
\begin{minipage}{2.5in}
\centering
\fbox{\includegraphics[height=1.6in,width=2.2in]
{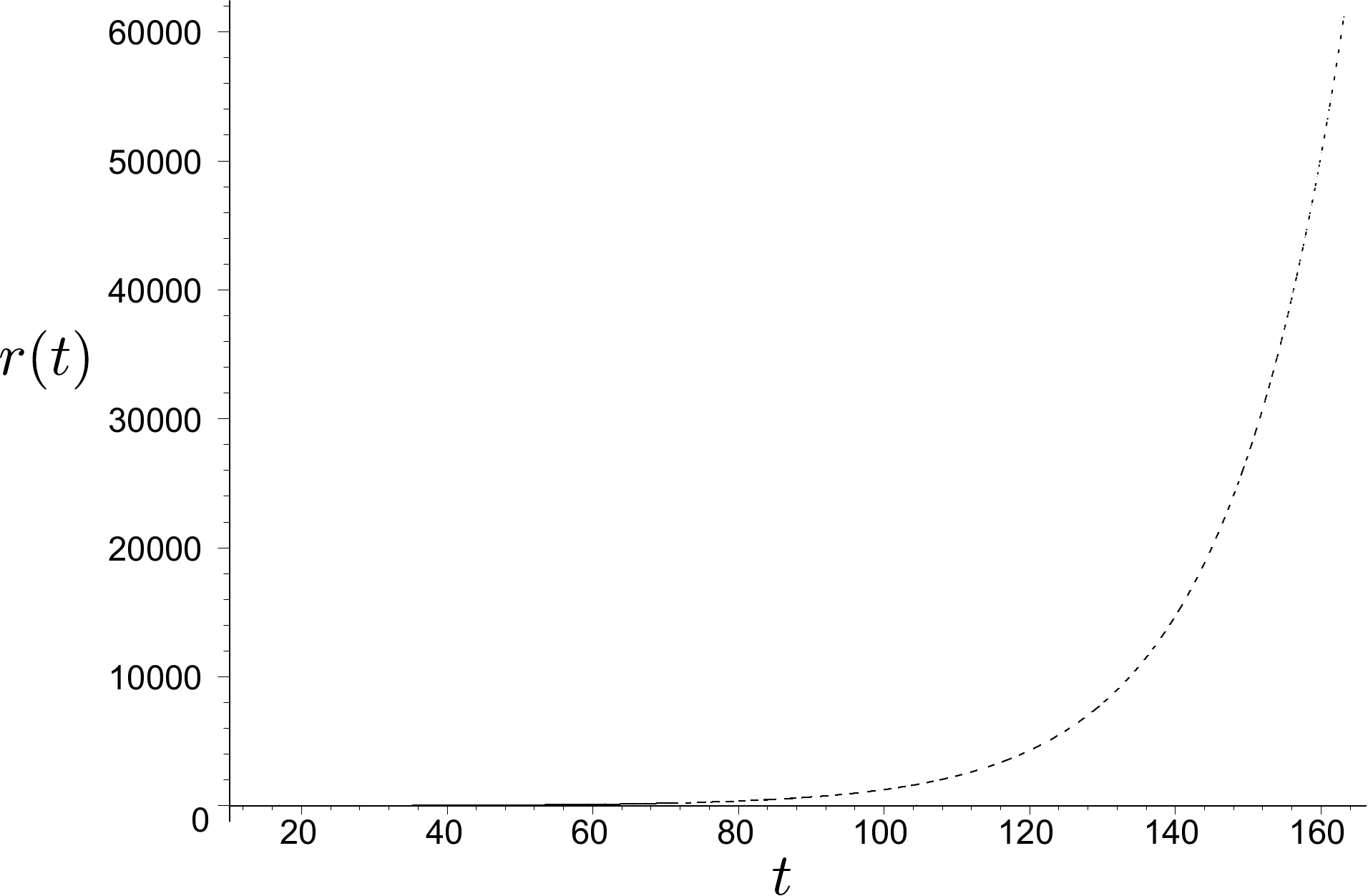}}
\end{minipage}
\\
(a) $r_i=5.575$ or $\bar{r}=6.34$ & 
(b) $r_i=5.58$
\\
\newline
\\
\begin{minipage}{2.5in}
\centering
\fbox{\includegraphics[height=1.6in,width=2.2in]
{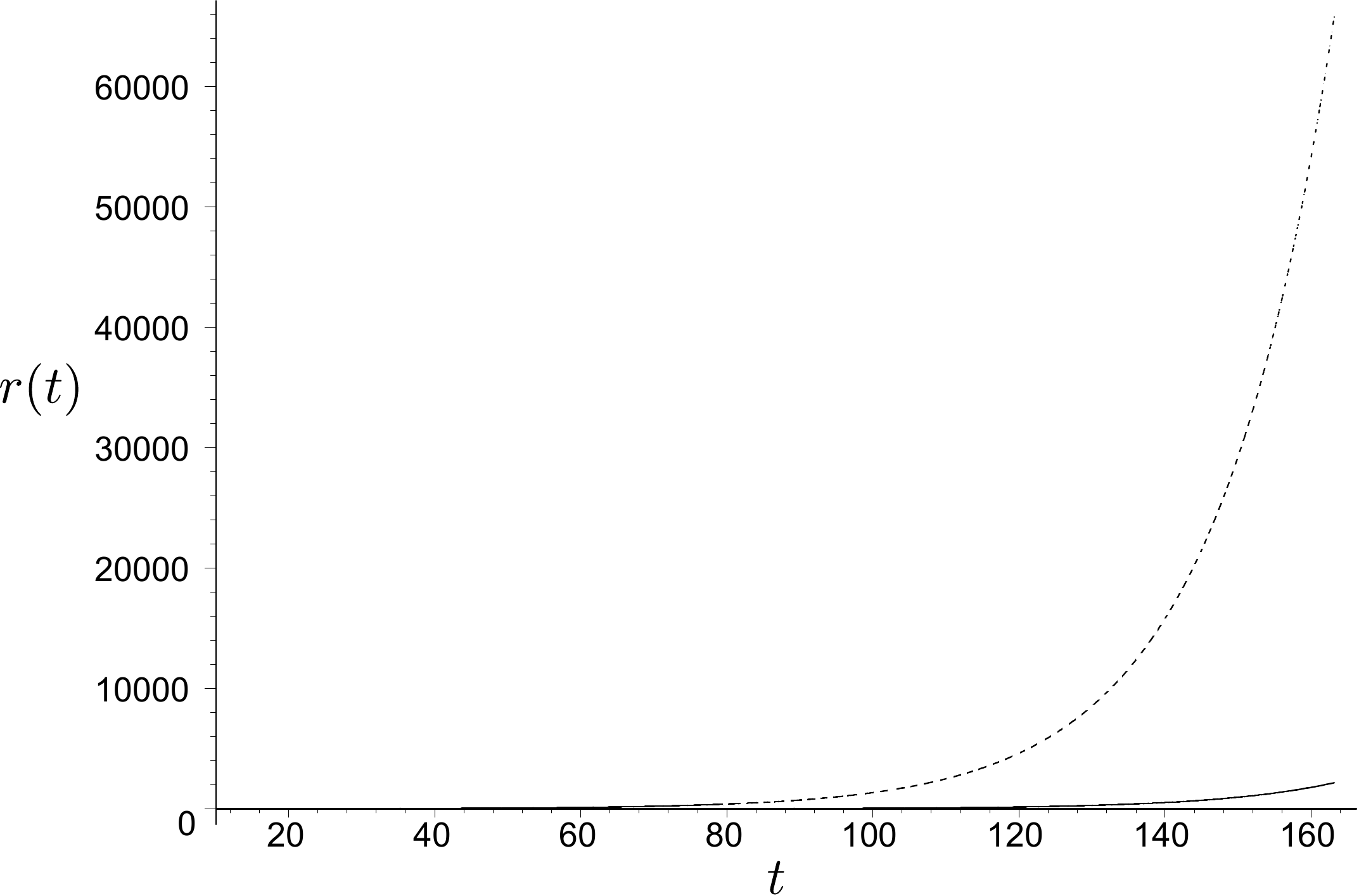}}
\end{minipage}
&
\begin{minipage}{2.5in}
\centering
\fbox{\includegraphics[height=1.6in,width=2.2in]
{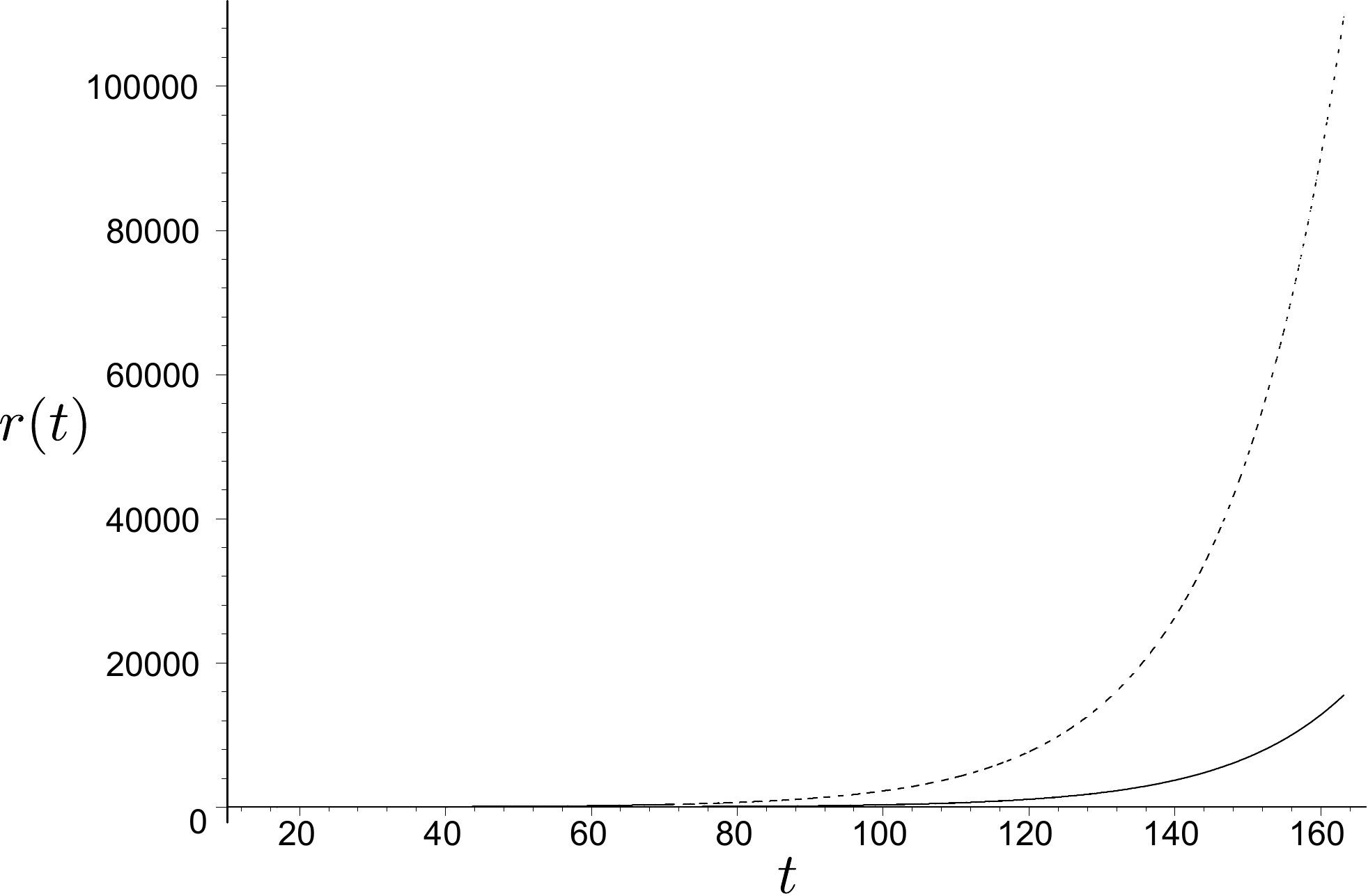}}
\end{minipage}
\\
(c) $r_i=6$ &
(d) $r_i=10$
\end{tabular}
\caption{Trajectories $r(t)$ (in Mpc versus Gyr) of a test particle
  launched at the present cosmic epoch $t=t_0$, at different initial
  radii $r_i$ from a central mass $m=10^{15}M_{\odot}$, such that its
  orbit is initially circular. The $\Lambda$CDM concordance model
  universal expansion law (\ref{eq:concordance}) is
  assumed.\label{fig:num1}}
\end{figure*}

\subsubsection{Numerical results}

As a check on our analytical results, and to include more accurately
the effect of the radius $r_S(t)$ being a function of cosmic time, we
perform a numerical integration of the Newtonian equation of motion
(\ref{eq:newtgeo}). In particular, we consider a test particle
launched at the present cosmic epoch $t=t_0$, at an initial radius
$r_i$ from the central mass $m$, such that its orbit is initially
circular. This corresponds to the boundary conditions $dr/dt =
d^2r/dt^2 = 0$ at $t=t_0$, and thus fixes the specific angular
momentum to be
\begin{equation}
L=\sqrt{mr_i + q(t_0)H^2(t_0) r_i^4}.
\label{eq:lmax}
\end{equation}
We take $m=10^{15}M_{\odot}$, which corresponds to a typical large cluster,
and assume a realistic expansion law $R(t)$, corresponding to the
standard spatially-flat concordance model, which is given by
\citep{GRbook}
\begin{equation}
a(t)=\left[\frac{(1-\Omega_{\Lambda,0})}{\Omega_{\Lambda,0}}
\sinh^2\left(\tfrac{3}{2}H_0\sqrt{\Omega_{\Lambda,0}}t\right)
\right]^{\frac{1}{3}},
\label{eq:concordance}
\end{equation}
where $a(t)=R(t)/R_0$ is the normalised scale factor,
$\Omega_{\Lambda,0} = 0.7$ and $H_0 = 70$ km s$^{-1}$ Mpc$^{-1}$.

It is found that there is a critical initial radius, $r_{i,c} \approx
5.575$ Mpc, above which the resulting particle trajectory expands with
the universe. Physically this means that a test particle in an initial
orbit of radius $r_i > r_{i,c}$ would ultimately be swept out by the
universal expansion. The resulting trajectory $r(t)$ for an initial
radius equal to the critical value is shown in Fig.~\ref{fig:num1}(a),
whereas the trajectories for larger values of $r_i$ are shown in the
other panels.  One sees that for the critical value, the radius of the
particle orbit oscillates about a fixed value $\bar{r}\approx 6.3$~Mpc
that is slightly larger than $r_i$, which itself lies at the minimum
of the oscillation. For larger values of $r_i$, the radius of the
resulting particle orbit grows significantly with cosmic time.
Nonetheless, in these latter cases, the growth of the particle orbit
lags behind the universal expansion. Although this `lag' becomes less
pronounced for larger values of $r_i$, the test particle never becomes
completely comoving owing to the presence of the central mass.

We note that \citet{price} has used a very similar geodesic equation
to (\ref{eq:newtgeo}) to analyse the motion of an electron orbiting a
central nucleus, with a slight modification to account for the
dominant electrostatic force in that problem.  An extension of this
work is presented in Appendix \ref{app:NLH2}.

\subsection{Minimum density of objects}

If we consider $r_S(t)$ as the maximum possible size of a massive
object at cosmic time $t$, and make the simplifying assumption that
objects are spherically-symmetric and have constant density, then the
condition (\ref{eq:maxstable}) shows that there exists a
time-dependent minimum density for objects, given by
\begin{equation}
\rho_{\text{min}}(t)=\frac{3m}{4\pi r_S^{\ 3}(t)} = -\frac{3q(t)H^2(t)}{\pi G},
\label{eq:rhomin}
\end{equation}
where in the last equality we have reinserted a factor of $G$ to
convert back to SI units. It should be noted that, since
(\ref{eq:maxstable}) is valid only for $q(t) < 0$ (accelerating
expansion), then this restriction also applies to
(\ref{eq:rhomin}). For $q(t) > 0$, $\rho_{\rm min}(t)$ vanishes.

We point out that in the special case of a de Sitter background, \citet{nowakowski1} calculate the minimum density for objects to remain gravitationally bound to be $\rho_{\rm min}=\Lambda/(4\pi G)$.  This is a factor of 4 smaller than our result for a de Sitter background.  It does however, correspond to what we would obtain by defining $\rho_{\rm min}(t)=3m/(4\pi r_F^{\ 3})$, with $r_F=(3m/\Lambda)^{1/3}$.  Since $r_F$ corresponds only to the {\itshape{theoretical}} maximum size of an object, and not the true maximum {\itshape{attainable}} size given by $r_S=(3m/(4\Lambda))^{1/3}$, we believe our estimate of the minimum density to be more realistic.

Using (\ref{eq:rhomin}) one finds that, at the present cosmic epoch, $\rho_{\rm min}(t_0) \sim
4 \times10^{-26}$ kg m$^{-3}$, which should be compared with the
present-day critical density $\rho_{\rm crit,0} \sim 9
\times10^{-27}$ kg m$^{-3}$ required for the universe to be
spatially-flat. Indeed, since at any cosmic epoch, 
$\rho_{\rm crit}(t) = 3H^2(t)/(8\pi G)$, one has simply
\begin{equation}
\frac{\rho_{\text{min}}(t)}{\rho_{\rm crit}(t)}= -8q(t),
\end{equation}
provided $q(t) < 0$. Alternatively, since the average matter density
$\rho_m(t) = \Omega_{\rm m}(t)\rho_{\rm crit}(t)$, the minimum
fractional density contrast $\delta_{\rm min}(t) \equiv [\rho_{\rm
    min}(t)-\rho_{\rm m}(t)]/\rho_{\rm m}(t)$ of a matter fluctuation
at cosmic time $t$ is
\begin{equation}
\delta_{\rm min}(t) = -\left[1+\frac{8q(t)}{\Omega_{\rm m}(t)}\right].
\end{equation}
Inserting the present-day values $\Omega_{\rm m,0} \approx 0.3$ and
$q_0 \approx -0.55$, one finds $\delta_{\rm min}(t_0) \approx 14$.

The above expressions and values are, of course, rather approximate.
In particular, our $r_S$ values have been calculated assuming a
central mass point rather than an extended mass and realistic
astrophysical objects do not have uniform densities. Nonetheless, the
above arguments can be taken as an indication of the
{\itshape{existence}} and rough order of magnitude of such a minimum
density.  Calculated accurately, this minimum density should
correspond to that required for any structure formation at all to
occur at a given cosmic time $t$ during an accelerated expansion phase
of the universe. We leave this as a topic for future research.

\subsection{Comparison with previous studies}\label{sec:comparison}

\citet{now3} and \citet{nowakowski2} have
previously used the Schwarzschild de Sitter metric (\ref{eq:SDS}) to
find expressions for (what we call) $r_F$ and $r_S$ in the special
case of a de Sitter background universe.

It is worth noting that they were able to use a more traditional
method to find $r_F$ and $r_S$, since the Schwarzschild de Sitter
metric is static. In such cases, one can combine the resulting
geodesic equations into an `energy' equation of the form
$\frac{1}{2}\dot{r}^2+U_{\text{eff}}(r)=E$, where $E$ is the
total energy of the particle per unit rest mass and one can read off
the effective potential $U_{\text{eff}}(r)$. For the Schwarzschild
de Sitter metric (\ref{eq:SDS}), this is given by
\begin{equation}
U_{\text{eff}}(r)=-\frac{m}{r}-\frac{1}{6}\Lambda r^2+\frac{L^2}{2r^2}-\frac{mL^2}{r^3}. \label{eq:effpot}
\end{equation}
The value of $r_F$ is then simply the location of the minimum of
$U_{\text{eff}}(r)$ with $L=0$, which immediately leads to an
expression for $r_F$ which agrees precisely with our result
(\ref{eq:accurateforce}) in the special case of a de Sitter
background. The value of $r_S$ is the location of the minimum of
$U_{\rm eff}(r)$ with $L=L_{\rm max}$, where $L_{\rm max}$ is maximum
value of $L$ for which $U_{\text{eff}}(r)$ possesses a minimum. In the
limits $L\gg m$ and $\Lambda\ll 1$ adopted by these authors, the
resulting expression for $r_S$ agrees with our result
(\ref{eq:maxstable}) in the de Sitter case.

This approach should be contrasted with the derivation in NLH1 of the
equation of motion (\ref{eq:fullgeo}). The latter is more complicated
owing to the dependence of the metric (\ref{eq:flat}) on the cosmic
time $t$, which prevents one arriving at an `energy'
equation. Nonetheless, we note that in the special case of a de Sitter
background, the equation of motion (\ref{eq:fullgeo}) reduces to
\begin{equation}
\ddot{r}=-\frac{m}{r^2}+\frac{1}{3}\Lambda r+\frac{L^2}{r^3}\left(1-\frac{3m}{r}\right).
\end{equation}
The first integral of this equation then immediately yields an
`energy' equation with the effective potential (\ref{eq:effpot}).

We further note that \citet{sussman} also obtained the factor
$r_F/r_S\approx 1.6$ for a de Sitter background, but individual values
of $r_F$ and $r_S$ are greater than ours by a factor of $\sim 10$ for
galaxies and $\sim 3$ for clusters.  Their approach deviates
significantly from our work and that of \citet{now3} and \citet{nowakowski2}, since they model galaxies and
clusters as dark matter haloes in equilibrium with a static
`$\Lambda$-field', rather than simply as point masses.  It is possible that, by taking the spatial extent of the central object into account, we may close this small discrepancy between our results, but we leave this as a topic for future research. We point out that Sussman \& Hernandez obtain the values of $r_F$ and $r_S$ by
numerically integrating two coupled differential equations, so no
analytical results are given in their work.

\section{Phantom energy and the `Big Rip'}\label{eq:bigrip}

We have thus far considered dark energy only in the form of a
cosmological constant, for which the equation-of-state parameter
$w=-1$. This is well supported by observations, but more generic dark
energy models allow for $w$ to differ from $-1$, and even to vary with
cosmic epoch. In any case, $w=-1$ is usually considered as the
lower-limit, since a dark energy `fluid' with $w < -1$ violates all
the relativistic energy conditions. Nonetheless, this has not deterred
some theorists from considering the cosmological consequences of a
dark energy component with a constant equation-of-state parameter $w <
-1$, which is often termed `phantom-energy' \citep{caldwell2}. We now
consider the effect on massive objects of a cosmological expansion
driven by phantom energy.

\subsection{The Big Rip}

As shown by \citet{caldwell}, the most profound consequence of phantom
energy is that it leads to a `Big Rip' in the far future, at which the
universe scale factor and Hubble parameter become singular.  

To explain this, for simplicity let us consider a spatially-flat
model, which is a good approximation to our universe. The Friedmann
equation governing the time evolution of the normalised scale factor
$a(t)$, assuming negligible radiation energy density, is
\begin{equation}
H^2(t)=H_0^{\ 2}\left[\Omega_{\rm m,0}a^{-3}(t)+(1-\Omega_{\rm m,0})a^{-3(1+w)}(t)\right],\label{eq:Friedmann}
\end{equation}
where $H(t)=a'(t)/a(t)$, $\Omega_{\rm m,0} \approx 0.3$ is the
present-day matter density parameter and the current value of the
Hubble parameter is taken to be $H_0\approx70$ km s$^{-1}$ Mpc$^{-1}$.

At late times the second term on the right-hand side, relating to dark
energy, dominates over the first, relating to matter.  Writing $1+w$
as $-|1+w|$, the Friedmann equation may be written as
$H^2(t)=H_0^{\ 2}(1-\Omega_{\rm m,0})a^{-3|1+w|}(t)$. Solving for
the normalised scale factor then gives
\begin{equation}
a(t)=\left[\frac{2}{3|1+w|\left(H_0\sqrt{1-\Omega_{\rm m,0}}(t_0-t)+\frac{2}{3|1+w|}\right)}\right]^{\frac{2}{3|1+w|}}.\label{eq:atsoln}
\end{equation}
One sees immediately that $a(t)$, and hence $H(t)$, diverges at a
finite time $t=t_{\rm rip}$ in the future that satisfies
\begin{equation}
t_{\rm rip}-t_0\approx \frac{2}{3|1+w|H_0\sqrt{1-\Omega_{\rm m,0}}}.\label{eq:trip}
\end{equation}
If $w=-\frac{3}{2}$, for example, the time remaining for the universe
before the Big Rip is $\sim 22$ Gyrs.

\subsection{Times at which objects become unbound}

Since $H(t)$ becomes infinite at $t=t_{\rm rip}$, then so does the
cosmological component of the radial force (\ref{eq:eq1again}) on a
test particle in the vicinity of a central mass. Moreover, since $q(t)
< 0$ for the accelerating expansion during phantom-energy dominance,
this force is directed ouwards, away from the central mass.  Thus, by the
time of the Big Rip, all massive objects will have been ripped apart.

We now consider the time before the Big Rip that this occurs for an
object of a mass $m$ and size $r$.  Note that we only perform a rough calculation due to our simplifying assumptions.  For example, we assume the mass to be concentrated at the centre of the object, as in a standard Newtonian framework.  We then specifically calculate the time at which a particle at the edge of the object becomes gravitationally unbound from it; that is when the inwards force component of magnitude
    $m/r^2$, due to the central mass, is balanced by the outwards
    component of magnitude $|q(t)H^2(t)r|$ due to the (accelerating)
    cosmological expansion.  Note that this is not the same as calculating the time at which the point mass itself is ripped apart, which we are unable to evaluate due to the singularity in our model at $r=2m$.  Nonetheless, it is the time at which the object ceases to exist in its current state and begins to be ripped apart, as layers of it gradually peel off as the cosmological force increasingly dominates over the gravitational force.  Since the particle is at a distance $r>>m$, the physical singularity in our model at $r=2m$ does not affect the result.

At late times, substituting $a(t)$ from (\ref{eq:atsoln}) into the
appropriate form of the Friedmann equation
$H^2(t)=H_0^2(1-\Omega_{\rm m,0})a^{-3|1+w|}(t)$, one finds
\begin{equation}
H^2(t)=\left(\frac{2}{3|1+w|\tilde{t}}\right)^2,
\end{equation}
where $\tilde{t}=t_{\rm rip}-t$ is the time before the Big Rip. It is
worth noting that this result is independent of both $H_0$ and
$\Omega_{\rm m,0}$.  Moreover, we see that at late times the deceleration
parameter tends to the constant value
\begin{equation}
q(t)\rightarrow-\frac{1}{2}|1+3w|.
\end{equation}
Thus the time before the Big Rip at which the total radial force
(\ref{eq:eq1again}) on a particle at the surface of an object of mass
$m$ and radius $r$ vanishes is given by
\begin{equation}
\tilde{t}\approx\frac{\sqrt{2|1+3w|}}{3|1+w|}\sqrt{\frac{r^3}{m}}.\label{eq:timebefore}
\end{equation}
Although this is only an approximate result, based on the Newtonian
force expression (\ref{eq:eq1again}), it remarkably matches Caldwell's
approximate expression precisely, despite the latter being derived in
a rather different manner based on energy arguments.  This could be merely a coincidence, but to determine the true extent to which our two approaches may in fact be parallel requires further thought.  We leave this for future research.

For $w=-3/2$, the times $\tilde{t}$ before the Big Rip that specific
objects of mass $m$ and size $r$ are expected to rip apart are given
in Table \ref{table:bigrip}. It is assumed (admittedly, rather
implausibly) that, as the universe evolves, the objects do not accrete
any additional mass, that their size remains fixed at the present-day
value until they are ripped apart, and that non-gravitational binding
forces are negligible.
\begin{table}
\begin{center}
\begin{tabular}{@{}lccc@{}}\hline
Object & $m$ & $r$ & $\tilde{t}$ \\ \hline
Galaxy cluster & $10^{15}$ M$_{\odot}$ & 5 Mpc & 9.3 Gyr  \\ 
Milky Way & $5.8 \times 10^{11}$ M$_{\odot}$ & 15 kpc & 69.9 Myr \\ 
Solar System & M$_{\odot}$ & $150\times 10^9$ m &3.6 months \\ 
Earth & $6\times 10^{24}$ kg & $6371\times 10^3$ m & 25.2 mins \\ 
\hline
\end{tabular}
\end{center}
\caption{The times $\tilde{t}$ before the Big Rip at which objects
  with a given mass $m$ and size $r$ are expected to be ripped apart,
  assuming a phantom energy equation-of-state parameter $w=-3/2$. In
  this case, the time until the Big Rip is $t_{\rm rip}-t_0 \approx
  22$ Gyr.\label{table:bigrip}}
\end{table}

\section{Conclusions}
\label{sec:conc}

In \citet{NandraPub1} we derived the metric for a point mass $m$
embedded in an expanding cosmological background, for both
spatially-flat and spatially-curved universes. We also derived the
corresponding invariant expression in each case for the force required
to keep a test particle at rest relative to the central mass. In this
paper, we have presented some important astrophysical consequences of
this work, focussing on the effect of an expanding universe on massive
objects on the scale of galaxies and clusters.

For a spatially-flat universe in the Newtonian limit, the force on a
test particle at a distance $r$ from a point mass $m$ consists of the
usual time-independent $1/r^2$ inwards component due to the central
mass and a time-dependent cosmological component proportional to $r$
that is directed outwards (inwards) when the expansion of the universe
is accelerating (decelerating). It is immediately clear, therefore,
that the effect of an expanding universe on a massive object is
profoundly different if the expansion is accelerating as opposed to
decelerating. In the latter case, the force on a constituent particle
of a galaxy or cluster is attractive for all values of $r$ and tends
gradually to zero as $r \to \infty$ (for any sensible radial density
profile), with no definite cut-off.  In the former case, however, the
force on a constituent particle (or equivalently its radial
acceleration) {\em vanishes} at a some {\em finite} radius $r_F$
(say), beyond which the net force becomes repulsive. This is also the
radius of the largest possible circular orbit about the central
mass. This suggests that an accelerating cosmological expansion should
set a {\em maximum size}, dependent on mass, for galaxies and
clusters.

Although our force expression accommodates a realistic dynamical
expanding background cosmology, it comes at the price of modelling the
central massive object as a point mass. In reality, it is more
appropriate to model galaxies or clusters as extended objects having
some density profile. We consider this in the special case of a
de Sitter background, which simplifies matters considerably.  In this
case, the expansion is always accelerating and the analysis can be
performed in a entirely static manner with the inclusion of a non-zero
cosmological constant.  Although the de Sitter background is not an
accurate representation of our universe, the standard cosmological
model is dominated by dark-energy in a form consistent with a simple
cosmological constant, which is driving the current observed
accelerated phase of expansion.  

We calculate the value of $r_F$ in this case for spherically-symmetric
matter distributions with an exponential and NFW radial density
profile, respectively, using Newtonian theory, linearised
general-relativity and the full non-linear Einstein field equations.
The three methods give indistinguishable results for objects with
masses and sizes typical of galaxies and clusters, as might be
expected. More surprising, perhaps, is that one can show analytically
that the location of $r=r_F$ is identical in the Newtonian and full
general-relativistic cases.  For clusters we find $r_F \sim
10.5$--12.0~Mpc and for galaxies we find $r_F \sim
1.06$--1.20~Mpc, which, as maximum sizes, are compatible with the sizes of $5$~Mpc and $0.26$~Mpc respectively.  In addition, for galaxies our values for $r_F$ agree well with the
typical radius at which the radial velocity field in the neighbourhood
of nearby galaxies vanishes, as inferred by \citet{peirani2, peirani1}
from recent observations of stellar velocities in such objects. In the
Newtonian and fully general-relativistic analyses, we include the
effects of pressure and derive the radial pressure profile $p(r)$ in
the central object. We find that $dp(r)/dr$ also vanishes precisely at
$r=r_F$, as does $p(r)$ itself according to our boundary condition.
In the Newtonian case, we obtain analytical results for pressure and
temperature profiles in the case of an NFW density profile.

The key difference between a general expression and the special case
of a de Sitter background is that, in the former, the cosmological
component of the radial force on a test particle is time-dependent.
This means that the radius $r_F$ at which the total radial force
becomes zero, which marks the cut-off point for the existence of
circular particle orbits, is also time-dependent.  In particular, for
the standard $\Lambda$CDM concordance cosmology, there is no finite
radius at which the total force vanishes, and hence no natural cut-off
for the size of massive objects, during the decelerating phase prior
to $z\approx 0.67$. Nonetheless, returning to our model of the central
massive object as a point mass, we find that at low redshifts, where
the dark energy component is dominant, the values of $r_F$ obtained do
not differ significantly from those obtained assuming a de Sitter
background.

For a general cosmological expansion, the radius $r_S$ of the largest stable circular orbit is also
time-dependent, but we find that at any cosmic epoch it is simply a
factor $4^{1/3} \approx 1.6$ smaller than $r_F$.  It is also perhaps a
more appropriate measure than $r_F$ for the maximum possible size of an
object of mass $m$.  Indeed it is encouraging that we find the factors $r_S/r_T$ at the present time for both galaxies and clusters to be close to 1; we obtain $\sim 2.8$ and $\sim 1.4$ respectively.  We also
obtain an approximate expression, dependent only on cosmological
parameters, for the minimum density of an object for it to remain
gravitationally bound at a given cosmic epoch.

Finally, we also consider the future effect on massive objects of
cosmological expansion driven by dark energy with an equation-of-state
parameter $w < -1$, termed phantom energy. In such a model the
universe is predicted to end in a `Big Rip' at a finite time in the
future at which the scale factor and the Hubble parameter become
singular. Using our radial force equation, we obtain an expression for
the time prior to the Big Rip that an object of a given mass and size
will become gravitationally unbound. Remarkably, this is found to
agree precisely with the approximate expression derived by Caldwell
using a rather different approach based on energy arguments.

\section*{Acknowledgements}

RN is supported by a Research Studentship from the Science and
Technology Facilities Council (STFC).

\appendix

\section[]{An atom's `all-or-nothing' behaviour}\label{app:c}

\begin{figure}
  \centering
\fbox{\includegraphics[height=2in,width=2.2in]{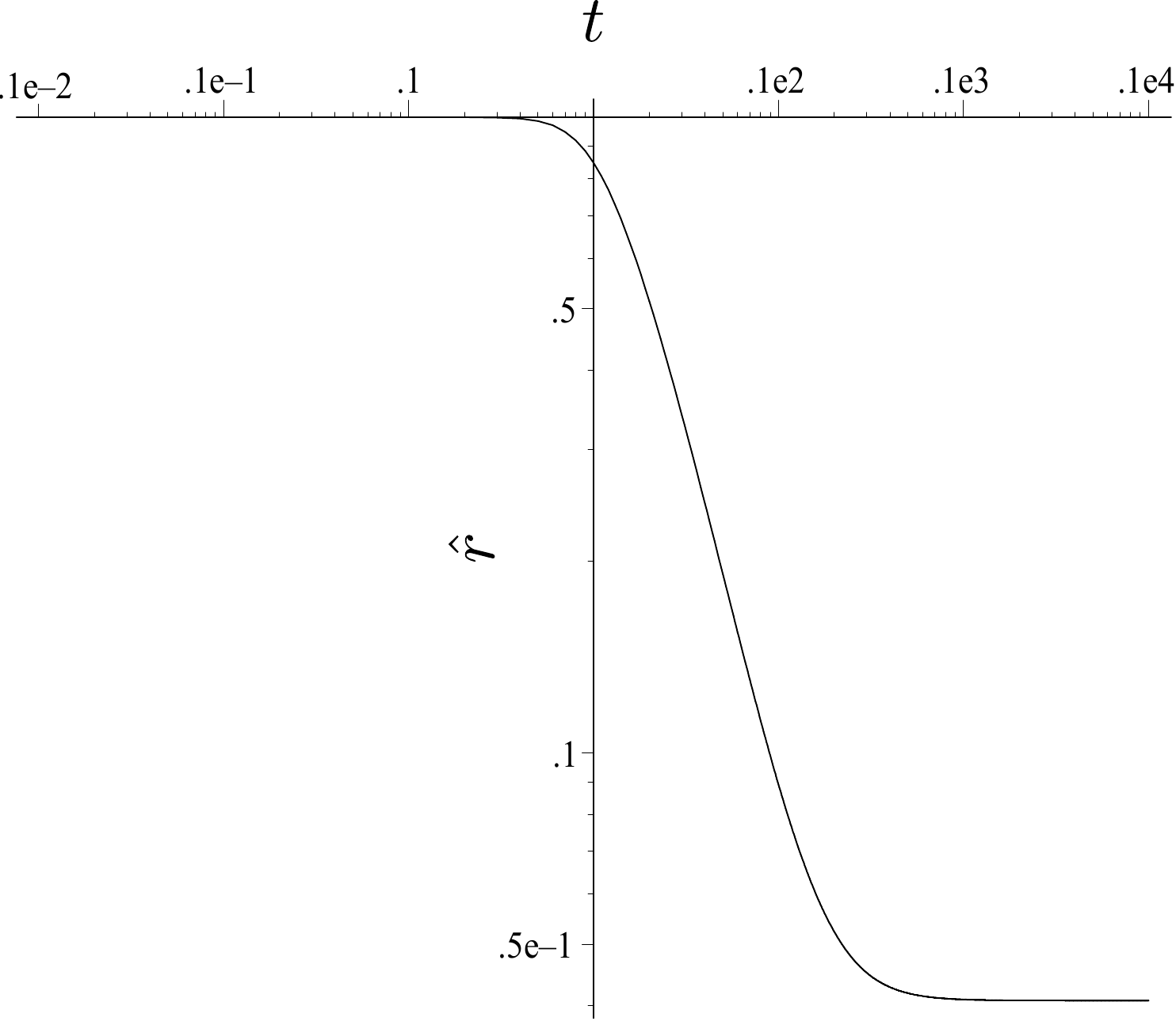}}
\caption{The cosmological radius $\hat{r}=r/a(t)$ of an electron orbit
  around a central nucleus for the critical value $L=3.46$ of the
  electron's specific orbital angular momentum.\label{fig:Price2}}
\end{figure}

\cite{price} has analysed the motion of an electron of mass $m_{\rm
  e}$ orbiting a central nucleus using the weak field, low velocity
geodesic equation (\ref{eq:newtgeo}).  For this model, he has shown
that it is appropriate simply to replace the $m/r^2$ gravitational
force in (\ref{eq:newtgeo}) with $C/r^2=QQ^{\prime}/(4\pi\epsilon_0
m_{\rm{e}}r^2)$, where $QQ^{\prime}$ is the product of the nuclear and
electron charges.  Price questions whether or not the electron
eventually takes part in the cosmological expansion; that is, does the
electron follow a trajectory of constant `physical' coordinate $r$ (no
atomic expansion) or of constant comoving coordinate $\hat{r}$ (full
cosmological expansion of the atom), or does the electron do something
`in-between'?  Using the model expansion law $a(t)=1+t^2\tanh(t)$,
where $a(t)=R(t)/R(t_0)$, and launching the test particle in a
momentarily circular orbit with $r=1$ (arbitrarily) at $t=0$, 
$(dr/dt)_{t=0}=(d^2r/dt^2)_{t=0}=0$ and thus
$C=L^2$, Price finds that there is a critical value of the electron's
specific orbital angular momentum $L \approx 3.46$ that determines an
`all-or-nothing' behaviour: that is a weakly bound atom $(L \leq
3.46)$ will eventually begin to stretch in size and participate wholly
in cosmological expansion (`all' behaviour), whereas a tightly bound
atom $(L > 3.46)$ will eventually ignore the cosmological expansion
and settle into a circular orbit of fixed `physical' radius (`nothing'
behaviour). We have been able to reproduce Price's results using his
expansion law and initial conditions. Fig.~\ref{fig:Price2} shows the
comoving radius of the electron orbit as a function of cosmic time $t$
for the critical value $L=3.46$, which shows an apparent plateau at
late times that Price interpreted as `all' behaviour.

\begin{figure}
  \centering
\fbox{\includegraphics[height=2in,width=2.5in]
{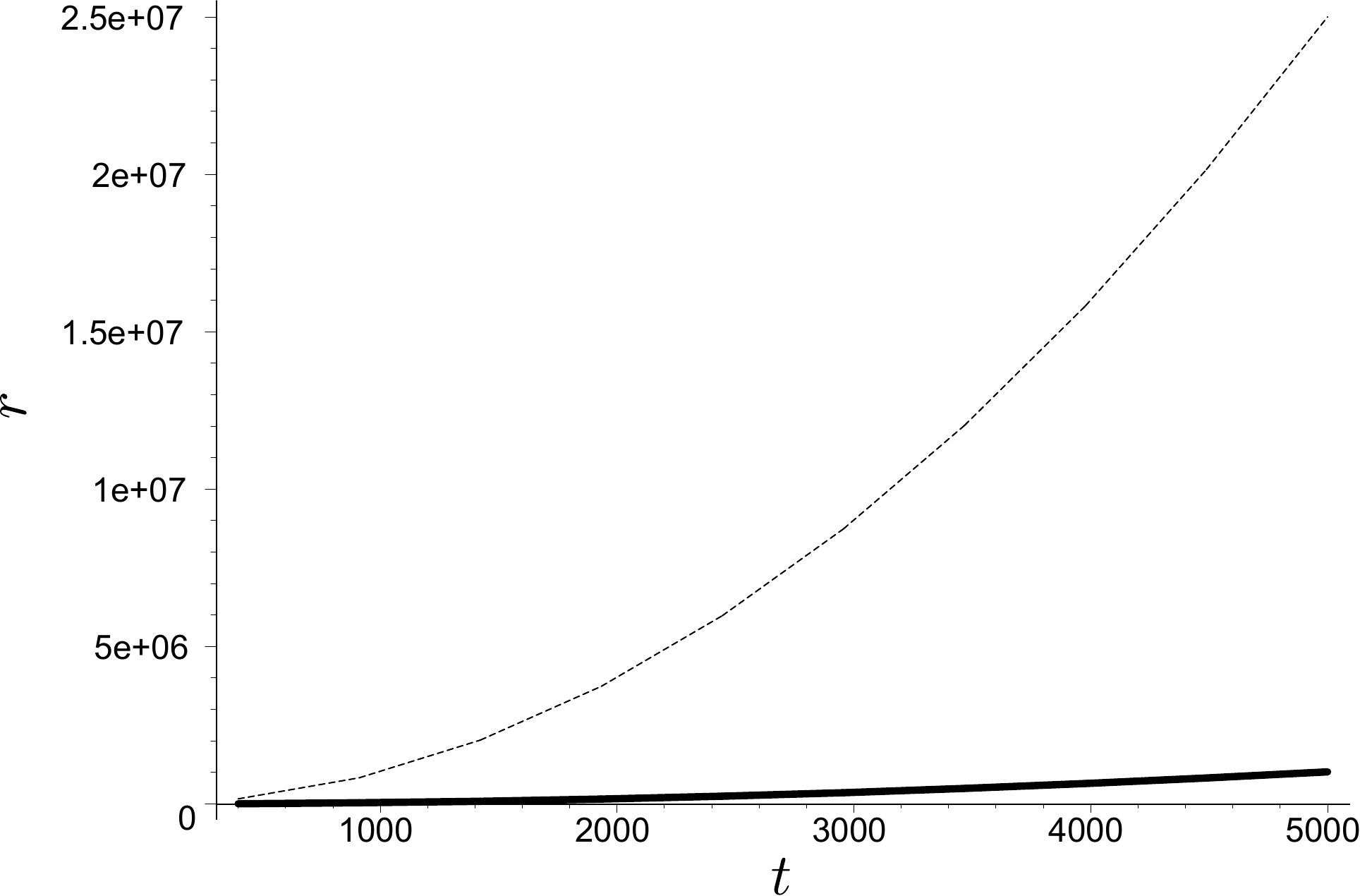}} \\[5mm]
\fbox{\includegraphics[height=2in,width=2.5in]
{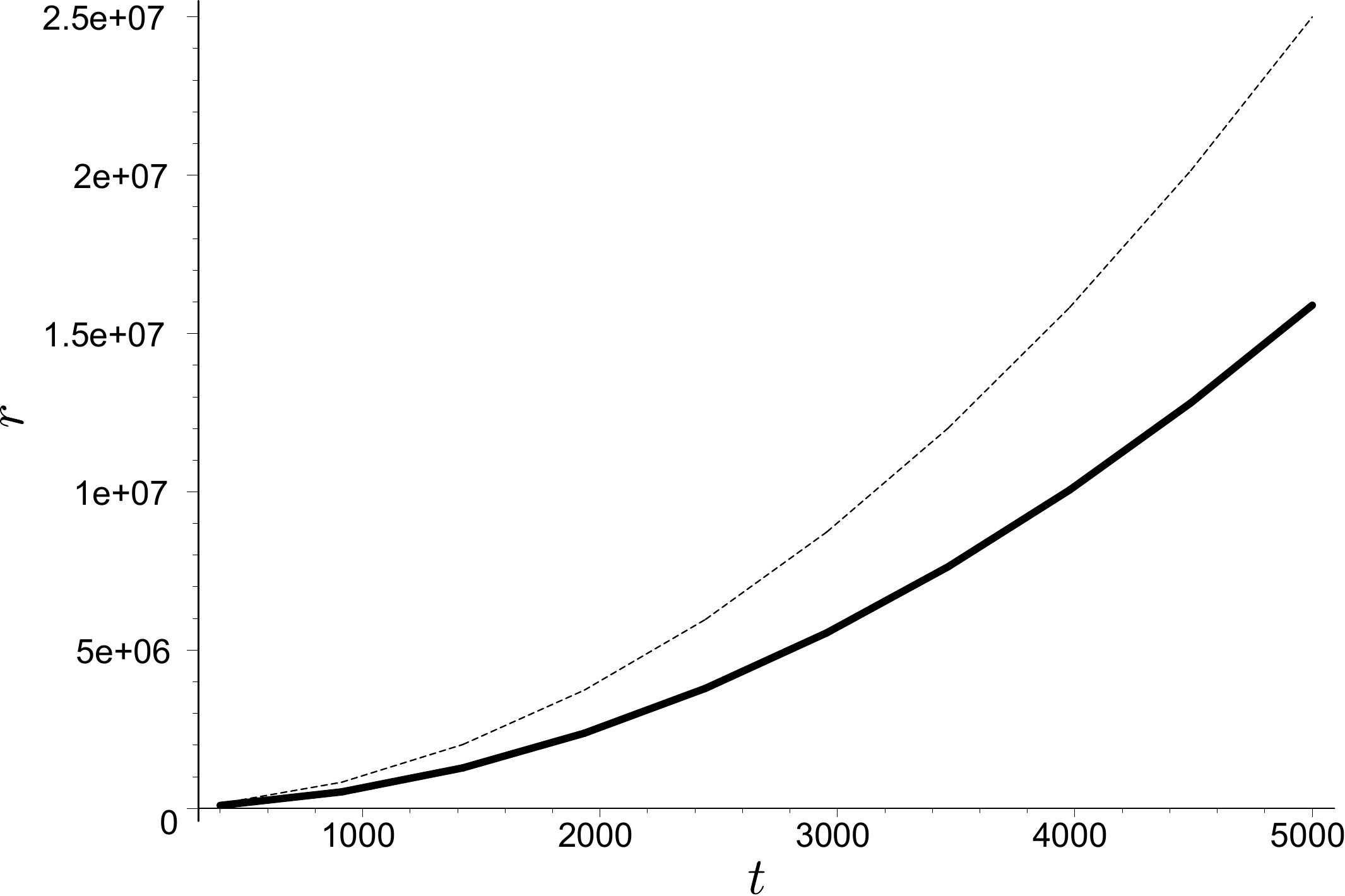}} \\[5mm]
\fbox{\includegraphics[height=2in,width=2.5in]
{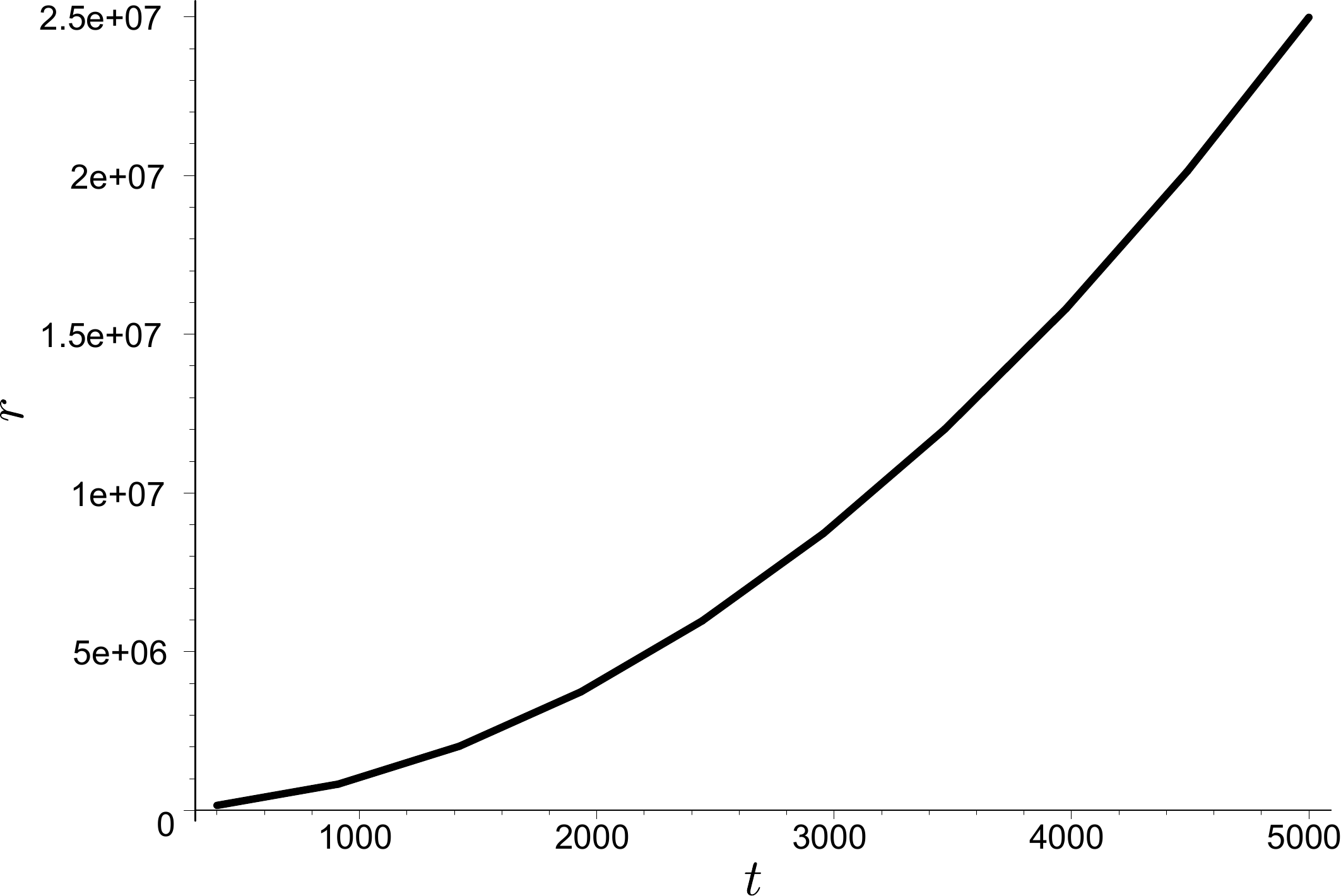}}
\caption{The `physical' radius $r$ (bold line) of an electron orbit
  around a central nucleus for different values of the electron's
  specific angular momentum: $L=3.46$ (top), $L=2.5$ (middle), $L=0.1$
  (bottom). The scale factor $a(t)$ is shown as the dashed
  line.\label{fig:Price3}}
\end{figure}

Closer inspection of the `all' behaviour reveals, however, a subtlety
that was not addressed previously.  The apparent plateau in
Fig.~\ref{fig:Price2} is not actually flat at all; rather $\hat{r}$
continues to decrease at late times but at a much slower rate than
previously.  For this value of $L$, the atom, in fact, exhibits only a
`something' behaviour. Fig.~\ref{fig:Price3} shows instead the
`physical' radius $r$ of the electron's orbit, together with the
universal scale factor $a(t)$. It then becomes clear that the atom
does indeed expand forever, but at a rate that increasingly lags
behind the cosmological expansion for $L=3.46$.  As we decrease $L$,
for even less tightly bound atoms, the amount by which the electron's
expansion lags behind the cosmological expansion decreases.
Eventually, for $L\leq0.1$, the two expand at the same rate and the
behaviour changes from `something' to `all'.

\bsp

\label{lastpage}

\end{document}